\documentclass{svjour3}
\RequirePackage{xcolor,graphicx}
\RequirePackage{natbib}
\usepackage[normalem]{ulem}
\usepackage{boxedminipage}
\usepackage{verbatimbox}
\usepackage[noend]{algorithmic}
\RequirePackage{multirow}
\RequirePackage{tikz,pgfplots}
\usetikzlibrary{shapes,arrows,positioning,matrix,backgrounds,fit}
\tikzstyle{marks} = [circle,fill=black,text=yellow,font=\small\sffamily\bfseries,inner sep=.1ex]
\RequirePackage{latexsym,amsmath,amssymb,bbm}
\RequirePackage{enumitem} \setlist{nosep}
\RequirePackage[colorlinks=true,allcolors=blue!60!gray]{hyperref}
\RequirePackage{soul}
\RequirePackage{colortbl}
\definecolor{qcolor}{RGB}{220,240,255}
\DeclareRobustCommand{\query}[1]{{\sethlcolor{qcolor}\sffamily\hl{#1}}}
\RequirePackage[textsize=small,color=red!20]{todonotes}
\DeclareMathOperator*{\argmax}{argmax}

\newsavebox{\fmbox}
\newenvironment{fmpage}[1]
{\begin{lrbox}{\fmbox}\begin{minipage}{#1}}
{\end{minipage}\end{lrbox}\fbox{\usebox{\fmbox}}}
\def\longbar{\relbar\joinrel\relbar\joinrel\relbar}

\AtBeginDocument{%
  \paperwidth=\dimexpr
    1in + \oddsidemargin
    + \textwidth
    + 1in + \oddsidemargin
  \relax
  \paperheight=\dimexpr
    1in + \topmargin
    + \headheight + \headsep
    + \textheight
    + 1in + \topmargin
  \relax
  \usepackage[pass]{geometry}\relax
}

\begin{document}

\title{Neural Architecture for Question Answering \\
  Using a Knowledge Graph and Web Corpus}
\titlerunning{Question Answering using Knowledge Graph and Corpus}

\author{Uma~Sawant \and Saurabh~Garg \and 
Soumen~Chakrabarti \and Ganesh~Ramakrishnan}

\institute{\relax}

\maketitle

\begin{abstract}
In Web search, entity-seeking queries often trigger a special Question Answering (QA) system.  It may use a parser to interpret the question to a structured query, execute that on a knowledge graph (KG), and return direct entity responses.  QA systems based on precise parsing tend to be brittle: minor syntax variations may dramatically change the response.  Moreover, KG coverage is patchy.  At the other extreme, a large corpus may provide broader coverage, but in an unstructured, unreliable form.  We present AQQUCN, a QA system that gracefully combines KG and corpus evidence.  AQQUCN accepts a broad spectrum of query syntax, between well-formed questions to short ``telegraphic'' keyword sequences.  In the face of inherent query ambiguities, AQQUCN aggregates signals from KGs and large corpora to directly rank KG entities, rather than commit to one semantic interpretation of the query.  AQQUCN models the ideal interpretation as an unobservable or latent variable.  Interpretations and candidate entity responses are scored as pairs, by combining signals from multiple convolutional networks that operate collectively on the query, KG and corpus.  On four public query workloads, amounting to over 8,000 queries with diverse query syntax, we see 5--16\% \emph{absolute} improvement in mean average precision (MAP), compared to the entity ranking performance of recent systems.  Our system is also competitive at entity set retrieval, almost doubling F1 scores for challenging short queries.
\end{abstract}

\section{Introduction}
\label{sec:Intro}

A large fraction of Web queries involve and seek entities \citep{LinPGKF2012ActiveObjects}.  Such queries may seek details of celebrities or movies (e.g., \query{kingsman release date}), historical events (e.g., \query{Who killed Gandhi?}), travel (e.g., \query{nearest airport to baikal lake}), and so on.  Queries that match certain patterns are handed off to specialized QA systems that directly return entity responses from a KG.  Sometimes, a semantic parse of the textual query is attempted \citep{BerantCFL2013SEMPRE,YihCHG2015STAGG} to translate it to a structured query over the KG, which is then executed to fetch a \emph{set} of response entities\footnote{These are known as KBQA or ``Knowledge Base Question Answering'' systems.}.  While providing precise answers if everything goes well, this approach to KG-driven QA is fraught with several difficulties.

\begin{itemize}
\item The input textual query may range from grammatically well-formed questions (e.g., \query{In which band did Jimmy Page perform before Led Zeppelin?}) to free-form ``tele\-gra\-phic'' keyword queries (e.g., \query{band jimmy page was in before led zeppelin}).  QA systems are often brittle with regard to input syntax, backing off if the input does not match specific syntactic patterns.

\item A curated, structured collection of facts in a KG reduces the QA task to ``compiling'' the textual query into a structured form which is directly executed on the KG. But KG coverage is always patchy --- with nodes and/or edges missing --- particularly when less popular entities are concerned.  For example, over 70\% of people in Freebase do not have a place of birth in Freebase~\citep{WestGMSGL2014KBCviaQA}.  If the types and relations expressed textually in the query cannot be mapped confidently to the KG, most QA systems back off.

\item Alternatively, one can extend IR-style text search by using an entity-annotated corpus of Web pages.  Any text snippet $s$ in the corpus which mentions entity $e$ and also matches the question $q$ well, can be considered supporting evidence for that entity $e$ to be the answer for~$q$.  However, such evidence from the Web corpus can be noisy due to incorrect entity linking of $s$ or $q$, and imperfect text matching between $q$ and~$s$.
\end{itemize}

\begin{figure}[t]
  \centering
  \begin{tikzpicture}[
      >=latex, font={\fontsize{8pt}{10}\selectfont},
      baseline=(current bounding box.center),
      typ/.style={draw=gray!50,fill=white,inner sep=.1mm, minimum height=2.3ex,},
      ent1/.style={draw=gray!50,fill=white,inner sep=.1mm,minimum height=2.3ex,},
      ent2/.style={draw=orange!40,fill=yellow!20,inner sep=.1mm, minimum height=2.3ex,},
      tn/.style={inner sep=.1mm, outer sep=.4mm, text depth=.1ex, minimum height=2.3ex,
        fill=white,dotted,draw=gray!60, dash pattern=on .2mm off .2mm},
      res/.style={dotted,draw=green!60!gray,line width=.4mm, dash pattern=on .4mm off .2mm},
      corp/.style={dotted,draw=blue!60!gray,line width=.4mm, dash pattern=on .4mm off .2mm},
    ]
    \node (music) [typ] {musical group};
    \node (lmusic) [above left=.2mm and 1mm of music,inner sep=0] {$t_2$};
    \draw (lmusic.south)--(music.west);
    \node (yard) [ent2,below left= of music] {The\_Yardbirds};
    \node (lyard) [below left=.2mm and 1mm of yard,inner sep=0] {$e_2$};
    \draw (lyard.north)--(yard.west);
    \draw (music.south west) edge [->] node[left]{\path{/type/object/type}}
    (yard.north east);
    \node (jimmy) [ent1,below right= of yard] {Jimmy\_Page};
    \node (ljimmy) [below left=.2mm and 1mm of jimmy,inner sep=0] {$e_1$};
    \draw (ljimmy.north)--(jimmy.west);
    \draw (yard.south east) edge [->] node(mgroup)[inner sep=0,
      outer sep=0]{\path{/music/musical_group/member}} (jimmy.north west);
    \scoped[on background layer]{\node (kg) [fit=(lmusic)(music)(lyard)
        (yard)(jimmy)(ljimmy)(mgroup),
        rounded corners, fill=gray!10, dashed, draw=gray!60,
        align=right, inner sep=.5mm] {};}
    \node [below right] at (kg.north west) {KG};
    \node [right=4mm of music, inner sep=0, outer sep=0] (tquery) {Query};
    \node (tband) [tn, below=1mm of tquery] {band};
    \node [tn, right=0 of tband] (tjim) {jimmy page};
    \node [tn, right=0 of tjim] (twas) {was in};
    \node [tn, right=0 of twas] (tbef) {before};
    \node [tn, right=0 of tbef] (tled) {led zeppelin};
    \path (music.east) edge[res,out=0,in=180] (tband);
    \path (jimmy) edge[res,out=90,in=270] (tjim);
    \path (mgroup.45) edge[res,out=90,in=270] (twas.south);
    \scoped[on background layer]{\node (query) [fit=(tquery)(tband)(tjim)(twas)
        (tbef)(tled), fill=blue!5, draw=blue!30,
        align=right, inner sep=.5mm] {};}
    \node (adot) [tn,below right=1mm and 10mm of mgroup,anchor=west] {\dots};
    \node (aabt) [tn,right=0 of adot] {about};  
    \node (apage) [tn,right=0 of aabt] {Page's};
    \node (awork) [tn,right=0 of apage] {work};
    \node (ain) [tn,right=0 of awork] {in};
    \node (athe) [tn,right=0 of ain] {the};
    \node (ayard) [tn,right=0 of athe] {Yardbirds};
    \node (aprior) [tn,below=.7mm of adot.south west, anchor=north west] {prior to};
    \node (aled) [tn,right=0 of aprior] {Led Zeppelin};
    \node (amdo) [tn,right=0 of aled] {\dots};
    \scoped[on background layer]{\node (snip1) [fit=(adot)(apage)(awork)(ain)
        (athe)(ayard)(aprior)(aled), fill=red!10, draw=orange!50,
        align=right, inner sep=.5mm] {};}
    \node [above left,inner sep=0,outer sep=0] at (snip1.south east) {Snippet};
    \path (aprior) edge[corp,out=90,in=270] (tbef);
    \path (aled) edge[corp,out=90,in=270] (tled);
    \path (yard.0) edge[corp,draw=red!70,out=0,in=175] (ayard.90);
  \end{tikzpicture}
  \caption{QA example using KG and corpus.
    Parts of the query have different \emph{roles} and match
    diverse artifacts in the KG and corpus, requiring a complex
    flow of evidence toward the correct response.}
  \label{fig:qaCorpusGraph}
\end{figure}
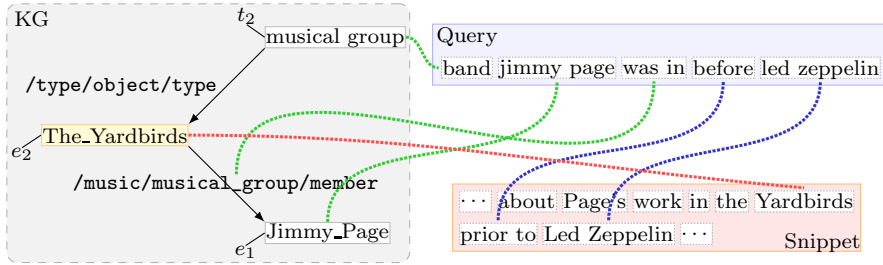

\paragraph{Example query and response:}
\figurename~\ref{fig:qaCorpusGraph} demonstrates the advantages and complexities of effective entity-level QA involving both KG and corpus. Tokens in the query have diverse, possibly overlapping \emph{roles}.  Specifically, a query span may hint at an entity, type or relation, or it can be used to match passages in the corpus.  Understanding the roles and disambiguating the hint to respective semantic nodes in the KG (wherever applicable) helps interpret that query. For example, the query $q=$ \query{band jimmy page was in before led zeppelin} has a reference to entities $e_1=\text{\path{Jimmy_Page}}$ and \path{Led_Zeppelin}.  The set of such entities, grounded in the query, will be called $\mathcal{E}_1$ with members~$e_1, e'_1$, etc.  Mentions in both the query and corpus documents are linked to entity nodes in the KG (e.g. \query{jimmy page}).  The target type of the query will be denoted~$t_2$ and a candidate answer entity will be denoted~$e_2$.  The set of all answer candidates will be called~$\mathcal{E}_2$, and the set of gold (ground truth) entities will be called~$\mathcal{E}^*_2$.  \query{Band} hints at $t_2 = \text{\path{musical_group}}$ of the expected answer entity $e_2 = \text{\path{The_Yardbirds}}$.  The (rather weak) hint \query{was in} hints at the relation $r=\text{\path{/music/musical_group/member}}$ connecting $e_1, e_2$. Thus, identifying $e_1$, $r$ and $t_2$ can lead us to many candidate $e_2$s, \text{\path{The_Yardbirds}} being one of them.  Yet, the query interpretation is not complete because an important token `before' is not considered.  If the KG does not have timestamps on membership, or the QA engine cannot do arithmetic with timestamps, passages in the corpus can still offer supplementary evidence by matching \query{before} with \query{prior to}, along with mentions of $e_1$ and~$e_2$.  Thus, using both KG and corpus allows combining structured and unstructured evidence to answer a query.

This example also serves to highlight our challenges.  The various curved, colored lines in \figurename~\ref{fig:qaCorpusGraph} map the query hints to the KG or the corpus evidence, either created during pre-processing stage (e.g. entity linking in the corpus) or at run-time (e.g. matching the target type $t_2 = \text{\path{musical_group}}$ to the query text \query{Band} through a type model).  The machine-learnt models which map the hints to KG entities, types or relations need to  handle a great deal of ambiguity, as a hint may match many correct and incorrect KG artifacts.  Thus, there may be multiple KG subgraphs and corpus text snippets, each appearing to support different correct or incorrect entity candidates.  There is thus a clear need for robust and seamless aggregation of supporting evidence across corpus and KG.

\paragraph{Our contributions:}
We present a new QA system, AQQUCN\footnote{Our system is named AQQUCN because it augments the AQQU system of \citet{BastH2015aqqu} with \underline{c}onvolutional \underline{n}etworks.}, with these salient features:
\begin{itemize}
\item AQQUCN is resilient to a spectrum of query styles, between syntactically well-formed questions to short `telegraphic' Web queries.  It does not attempt a grammar-based parse of the query.
\item AQQUCN uses KG and corpus signals in conjunction to score responses.  Rather than a single comparison network between query and corpus \citep{SeverynM2015SiamConvNet,BahdanauCB2014Attention}, AQQUCN uses a heterogeneous network architecture tailored to structural properties of queries.
\item Instead of choosing one structured interpretation and executing it on the KG to get a response \emph{set}, AQQUCN is capable of directly ranking entities based on evidence pooled over multiple structured interpretations.
\end{itemize}
We review related work in Section~\ref{sec:Rel}.  In Section~\ref{sec:OurSystem} we give an overview of AQQUCN, and in Section~\ref{sec:AqqucnModules} we describe all the modules in detail.  In Section~\ref{sec:results} we evaluate AQQUCN against recent competitive baseline systems.  Our code with relevant data will be made available \citep{CsawProject}.
\nocite{SawantCR2018aqqucn1}

\section{Related work}
\label{sec:Rel}

Recent QA systems are the result of convergence between several communities: Information Retrieval (IR), NLP, machine learning, and neural networks.

\subsection{Corpus-oriented entity search}
\label{sec:Rel:CorpusQA}

Early work in the IR community focused on corpus-driven QA in the TREC-QA track \citep{Wang2006QAsurvey, Cardie2012NLPCourse}.  The Web and IR community has traditionally assumed a free-form query that is often `telegraphic'.  Web search queries being far more noisy, the goal of structure discovery is more modest.  Indeed, in expert search, one of the earliest forms of corpus-based entity search focused on finding experts (people) in a given field, query structure discovery was given no importance.  State-of-the-art expert search systems \citep{BalogAdR2009LanguageModelExpert, MacdonaldO2011RankingAggregates, PetkovaBC2007NamedEntityProximity} collect text snippets (or documents) containing query words, match each snippet evidence to an expert (e.g. using the signal that the expert's name is mentioned in the snippet); and aggregate such evidence snippets to rank the experts.  Generative language models \citep{BalogAR2006ExpertSearch}, proximity based kernels \citep{PetkovaBC2007NamedEntityProximity} and feature-based supervised discriminative learning \citep{FangSM2010DiscriminativeExpertSearch} were evaluated to score the evidence match.  Conversely, document retrieval can be improved by expanding the query with entity features \citep{DaltonDA2014EQFE}.

\subsection{Entity search from knowledge graphs (KGQA/KBQA)}
\label{sec:Rel:KGQA}

As the information extraction and NLP communities developed more tools for annotating corpus spans with named entity (NE) types \citep{LingW2012FineType} and canonical entity IDs from KGs \citep{GaneaH2017DeepNed}, corpus-based techniques were refined to match answer types \citep{MurdockKWFFGZK2012TypeCoercion}.  With support from large KGs like Wikipedia and Freebase, the NLP community developed semantic parsers \citep{BerantCFL2013SEMPRE, YaoVD2014Jacana, YihCHG2015STAGG, KasneciSIRW2008NAGA, PoundHIW2012InterpretQuery, YahyaBERTW2012DEANNA, KwiatkowskiCAZ2013SemPar} that translated natural language queries to a target graph query language similar to SPARQL.  These approaches typically assume that question utterances are grammatically well-formed, from which precise clause structure, ground constants, variables, and connective relations can be inferred via semantic parsing.  A similar system called AQQU \citep{BastH2015aqqu} emerged from the IR community.  We base our system on it, so we will describe it separately in Section~\ref{sec:AqquReview}.  Such approaches are often correlated with the assumption that all usable knowledge has been curated into the KG.  The query is first translated to a structured form, which is then executed on the KG.

\begin{figure}[ht]
\centering
\begin{tikzpicture}
  \tikzset{factor/.style={outer sep=0, inner sep=0}}
  \node [draw, circle,
  label={[align=center]above:Query segmentation}] (z) {$z$} ;
  \node [factor, below left=10mm and 7mm of z,
    label={[align=right, text width=17mm]left:Potential between
      relation hint and $r$}] (zr) {$\rule{1ex}{1ex}$} ;
  \node [factor, left=25mm of zr,
    label={[align=right, text width=20mm]north west:Potential\\
      between type hint and $t_2$}] (zt2) {$\rule{1ex}{1ex}$} ;
  \node [factor, below right=10mm and 10mm of z,
    label={[align=right, text width=20mm]left:Entity
      linker score}] (ze1) {$\rule{1ex}{1ex}$} ;
  \node [factor, right=24mm of ze1,
    label={[align=left, text width=20mm]north east:Default\\
      potential for not interpreting some query words wrt KG}]
  (zs) {$\rule{1ex}{1ex}$} ;
  \node [circle, draw, below left=10mm and 5mm of zt2] (t2) {$t_2$} ;
  \node [circle, draw, below left=10mm and 0mm of zr] (r) {$r$} ;
  \node [circle, draw, below right=10mm and 0mm of ze1] (e1) {$e_1$} ;
  \node [circle, draw, below right=10mm and 5mm of zs] (s) {$s$} ;
  \node [factor, below right=10mm and 5mm of r,
    label={[align=right, text width=20mm]left:Potential for belief
      that $(e_1, r, e_2) \in \text{KG}$}] (pr) {$\rule{1ex}{1ex}$} ;
  \node [factor, below right=10mm and 5mm of e1,
    label={[fill=yellow!10, align=left, text width=25mm]south east:Presence
    and scores of corpus snippets mentioning $s, e_1, e_2$ and expressing\\
      relation~$r$}] (pse1e2) {$\rule{1ex}{1ex}$} ;
  \node [factor, below left=10mm and 19mm of r,
    label={[label distance=0, align=right,
        text width=20mm]south west:Potential for
      belief that $e_2 \in t_2$}] (pet) {$\rule{1ex}{1ex}$} ;
  \node [draw, circle, fill=gray!30,
    below right=10mm and 5mm of pr,
    label={[align=center]below:Fix/`observe'
      candidate\\ $e_2$ and score it}] (e2) {$e_2$} ;
  \path [draw] (z) -- (zt2) ;
  \path [draw] (z) -- (zr) ;
  \path [draw] (z) -- (ze1) ;
  \path [draw] (z) -- (zs) ;
  \path [draw] (zt2) -- (t2) ;
  \path [draw] (zr) -- (r) ;
  \path [draw] (ze1) -- (e1) ;
  \path [draw] (zs) -- (s) ;
  \path [draw] (t2) -- (pet) ;
  \path [draw] (r) -- (pr) ;
  \path [draw] (e1) -- (pr) ;
  \path [draw, dashed] (r) -- (pse1e2) ;
  \path [draw, dashed] (e1) -- (pse1e2) ;
  \path [draw] (s) -- (pse1e2) ;
  \path [draw] (pet) -- (e2) ;
  \path [draw] (pr) -- (e2) ;
  \path [draw] (pse1e2) -- (e2) ;
\end{tikzpicture}
\caption{The graphical model used by \citet{JoshiSC2014KgCorpusJoint} to score each candidate entity~$e_2$ in turn by enumerating over all other variables (nodes), which are unobserved or `latent'.  Factors nodes are denoted by~\rule{1ex}{1ex}.  (Although random variables in nodes are conventionally written in uppercase, we use lowercase to avoid confusion with \emph{sets} $\mathcal{E}_1, \mathcal{E}_2$, etc.)}
\label{fig:JoshiGraphModel}
\end{figure}
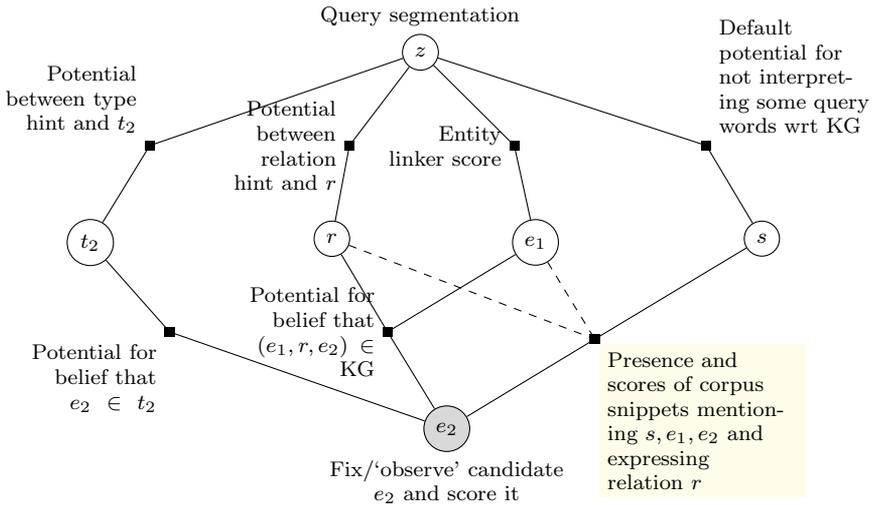

\subsection{Combining corpus and KG}
\label{sec:Rel:CorpusKGQA}

There is increasing interest in combining corpus and KG for QA.  AQQUCN is related to the single-relation QA system described by \citet{JoshiSC2014KgCorpusJoint}.  Unlike AQQUCN, they attempted an explicit 4-way segmentation of the query to identify the mention spans $\tilde{e_1}$ that mark grounded entities $e_1$, mention span\footnote{Hint of relation $r$ might be distributed among multiple disjoint spans, but this is not a serious problem for our proposed system because we allow spans with multiple roles.} $\tilde{r}$ that marks a hint to relation~$r$, span $\tilde{t_2}$ that marks a hint to the target type~$t_2$, and the remaining tokens $\tilde{s}$ are designated as \emph{selectors} that are meant to keyword-match corpus snippets.  This overall query segmentation is denoted~$z$.  The segmentation $z$ guides their system to propose \emph{structured} KG artifacts $e_1, r, t_2$ corresponding to query spans $\tilde{e_1}, \tilde{r}, \tilde{t_2}$.  The candidate response entity $e_2$ is then scored over admissible values of all latent variables (see \figurename~\ref{fig:JoshiGraphModel}).  KG and corpus signals are unified as various factors in the model.  We address two limitations in this system.  First, we do not attempt a hard query segmentation.  Second, we replace the traditional discrete language models that inform the factor potentials with continuous neural counterparts.

More work in this vein followed rapidly.  \citet{XuRFHZ2016FbCorpusQA} presented a KGQA system with a corpus-based postprocessing pruning stage that removes candidates with weak corpus support.  A more symmetric architecture called Text2KB was proposed by \citet{SavenkovA2016KbCorpusQa}.  A KGQA system as in Section~\ref{sec:Rel:KGQA} collects candidate answers.  A corpus search collects snippets from top-ranking documents and annotates \citep{GlobersonLCSRP2016MultiFocal, GaneaH2017DeepNed} them with KG entities.  For each candidate in the union, features are collected from both KG and corpus snippets to rank them.  In a similar spirit, \citet{XiongCL2017WordEntityDuet} propose AttR-Duet.  Both the query and corpus passage are represented as bags of words as well as entities.  Definition and mention texts in the KG and corpus are used to bridge between the space of entities and words, and four families of similarities ($\{\text{query}, \text{passage}\} \times \{\text{word}, \text{entity}\}$) are defined.  These are then combined using a neural network.  Reminiscent of how \citet{JoshiSC2014KgCorpusJoint} incorporated a corpus-based factor/potential into a graphical model (\figurename~\ref{fig:JoshiGraphModel}), \citet{BastB2017QLever} proposed QLever, which extended \citep{Chakrabarti2010WSDM} SPARQL with predicates over an entity-annotated corpus.  The primary focus of QLever is on high performance in the face of query clauses spanning KG and corpus indices, not ranking accuracy per se.

\subsection{Complex QA using neural techniques}
\label{sec:Rel:NeuralQA}

Early improvements to QA systems resulted from replacing discrete word matching and scoring with word vector counterparts.  In corpus-based QA, \citet{bordes2014question} modeled queries and passages as bags of words and simply added up their word embeddings to represent and compare them. \citet{yang2014joint} used embeddings to translate questions into relational predicates.  More refined sentence/query embeddings have been created \citep{SeverynM2015SiamConvNet, iyyer2014neural} via recurrent networks (RNNs) and convolutional  networks (CNNs), but usually applied to syntax-rich, well-formed questions. \citet{Dong:2015} obtained better accuracy than \citet{bordes2014question} by replacing the aggregated word vector query representation with multiple parallel CNNs for extracting deep representations for relation $r$ between query entity (i.e. the entity mentioned in the query) and answer entity, answer type and the KG neighborhood of the query entity.  While the above works dealt with entity retrieval, CNNs have also been actively explored in query-document matching for complex answer retrieval \citep{hui2017pacrr, hui2018co, macavaney2018characterizing}.  EviNets \citep{SavenkovA2017EviNets} embeds the query and evidence passage as average word vectors.  Then it collects various aggregates \citep{MacdonaldO2011RankingAggregates} of vector match scores, which are combined using a trained pooling network.  None of these neural systems seek a structural understanding of the query and how its parts relate differentially to the KG and corpus.  Recently, neural learning techniques are being used to translate very complex queries \citep{SahaPKKC2018CSQA} like \query{how many countries have more rivers than Brazil} into multi-layer expression graphs or multi-step imperative programs \citep{AndreasRDK2016ComposeNeuralQA, DongL2016lang2logic, MillerFDKBW2016KVMnet, ZhongXS2017seq2sql, ReedF2015NPI, LiangBLFL2016NSM}.  The extreme complexity of the  reinforcement learning formulations needed have, thus far, precluded scaling them to Web-scale corpora and noisy query and corpus annotation tools.

\section{AQQUCN overview}
\label{sec:OurSystem}

AQQUCN, a system we have built by extending AQQU \citep{BastH2015aqqu}, implements all the inference pathways shown in \figurename~\ref{fig:qaCorpusGraph}.  Unlike AQQU and some other systems, the end goal of AQQUCN is not to ``compile'' the input into a structured query to execute on the KG, because broken/missing input syntax can make this attempt fail in brittle ways. Instead, the end goal of AQQUCN is to directly create a \emph{ranking} over entities using KG and corpus, using interpretations as latent variables.  We first review AQQU briefly, and then describe the three stages of AQQUCN in the rest of this section.

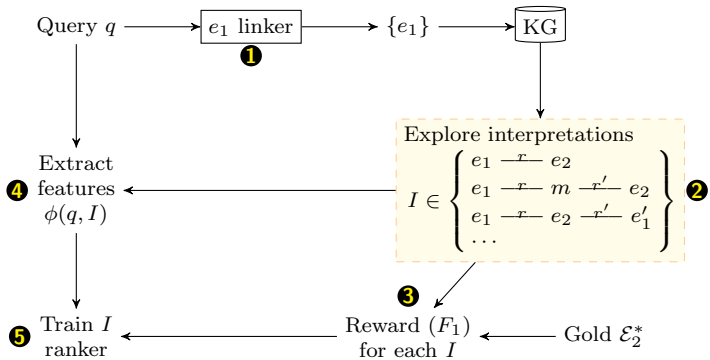
\begin{figure}[ht]
\centering
\begin{tikzpicture}
  \tikzset{>=stealth'}
  \node (query) {Query~$q$} ;
  \node [draw] (e1link) [right=of query] {$e_1$~linker} ;
  \node [marks, below=.1mm of e1link] {1};
  \path [draw, ->] (query) -- (e1link) ;
  \node (e1s) [right=of e1link] {$\{e_1\}$} ;
  \path [draw, ->] (e1link) -- (e1s) ;
  \node (kg) [cylinder, draw, right=of e1s,
    shape border rotate=90, shape aspect=0.1] {KG} ;
  \path [draw, ->] (e1s) -- (kg) ;
  \node (interps) [align=left, below=of kg, draw=orange!40,
  	dashed, fill=yellow!10] {
    Explore interpretations\\
    $\begin{aligned}
      I \in \left\{
      \begin{array}{l}
        e_1 \stackrel{r}{\longbar} e_2 \\
        e_1 \stackrel{r}{\longbar} m \stackrel{r'}{\longbar} e_2 \\
        e_1 \stackrel{r}{\longbar} e_2 \stackrel{r'}{\longbar} e'_1 \\
        \cdots
      \end{array}
      \right\}
    \end{aligned}$
  } ;
  \node [marks, right=.1mm of interps] {2} ;
  \path [draw, ->] (kg) -- (interps) ;
  \node (phi) [align=center] at (interps -| query) {
    Extract\\ features\\ $\phi(q, I)$};
  \node [marks, left=.1mm of phi] {4} ;
  \path [draw, ->] (query) -- (phi) ;
  \path [draw, ->] (interps) -- (phi) ;
  \node (ranker) [below=10mm of phi, align=center] {
  	Train~$I$\\ ranker} ;
  \node [marks, left=.1mm of ranker] {5} ;
  \node (loss) [align=center] at (ranker -| e1s) {
    Reward ($F_1$)\\ for each~$I$};
  \node [marks, above=.1mm of loss] {3} ;
  \node (e2star) [right=of loss] {Gold~$\mathcal{E}_2^*$} ;
  \path [draw, ->] (interps) -- (loss) ;
  \path [draw, ->] (e2star) -- (loss) ;
  \path [draw, ->] (loss) -- (ranker) ;
  \path [draw, ->] (phi) -- (ranker) ;
\end{tikzpicture}
\caption{Simplified sketch of AQQU \citep{BastH2015aqqu} --- training the interpretation ranker (follow the numbers).  The $e_1$ linker may shortlist alternative candidates, or candidates for multiple grounded entities~$e_1, e'_1$, etc.  Exploration around these produce interpretations $I$.  Each $I$ is associated with a system output $\mathcal{E}_2$ of a set of answer entities.  This is compared against the gold $\mathcal{E}^*_2$ to compute the reward~$F_1$, which guides the interpretation ranker.}
\label{fig:AQQU}
\end{figure}

\subsection{AQQU review}
\label{sec:AqquReview}

AQQU interprets the input question with reference to three possible ``query templates'', each having a direct translation to a SPARQL query.  First, grounded entities $e_1$ in the query are identified.  Then, a guided expansion in the KG locates candidates $e_2$, resulting in structured interpretations $I$ that may take forms such as these one- and two-hop queries:
\begin{itemize}
\item $e_1 \stackrel{r}{\longbar} e_2$.
\item $e_1 \stackrel{r}{\longbar} m
\stackrel{r'}{\longbar} e_2$, where $m$ is a mediator `entity' often representing a ternary relation.
\end{itemize}
AQQU then extracts features from question $q$, interpretation $I$ (which could have a few distinct structures as above), and all the candidate $e_2$s together.  These features are used in logistic regression or random forests to score each interpretation.  The best interpretation (with unbounded placeholders for $e_2$ and $m$, if applicable) is then executed on the KG.  During training, the gold interpretation is not known, but executing each interpretation gives a system response set $\hat{\mathcal{E}}_2$ that can be compared against the gold $\mathcal{E}^*_2$ to get an F1 score.  This helps AQQU train logistic regression or random forests to score better interpretations higher than worse ones.  The unit of scoring and ranking is a single interpretation, not an entity, nor a joint space of interpretation and entity, as is the case in AQQUCN.  Moreover, AQQU does not use corpus signals.

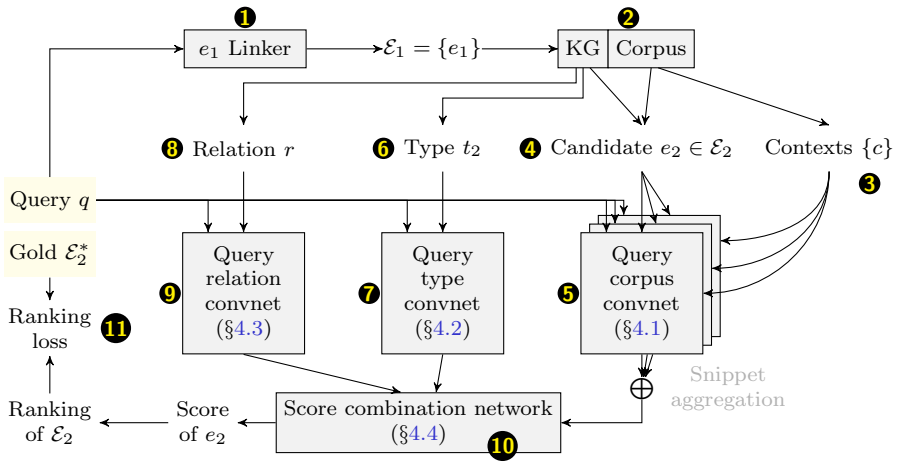
\begin{figure}[ht]
  \centering 
  \begin{tikzpicture}
    \tikzset{>=stealth'}
    \node [draw, fill=gray!10, minimum size=16mm,
      align=center] (qcn0) at (10mm, 45mm) {} ;
    \node [draw, fill=gray!10, align=center,
      minimum size=16mm, below left=-15mm and -15mm of qcn0]
    (qcn1) {} ;
    \node [draw, fill=gray!10, align=center,
      minimum size=16mm, below left=-15mm and -15mm of qcn1]
    (qcn2) {Query\\ corpus\\ convnet\\ (\S\ref{sec:qcn})} ;
    \node [marks, left=.1mm of qcn2] {5} ;
    \node [inner sep=0, below=3mm of qcn2] (agg) {$\bigoplus$} ;
    \draw [->] (qcn2.270) to (agg) ;
    \draw [->] (qcn2.275) to (agg) ;
    \draw [->] (qcn2.280) to (agg) ;
    \node [right=1mm of agg, align=center]
    (aggdesc) {\color{gray!60}Snippet\\ \color{gray!60} aggregation} ;
    \node [draw, fill=gray!10, minimum size=16mm,
      align=center, left=10mm of qcn2] (qtn) {Query\\ type\\ convnet\\
      (\S\ref{sec:qtn})} ; 
    \node [draw, fill=gray!10, minimum size=16mm,
      align=center, left=10mm of qtn] (qrn) {Query\\ relation\\ convnet\\
      (\S\ref{sec:qrn})} ; 
    \node [draw, fill=gray!10, minimum height=5mm, text depth=.25ex,
      outer sep=0, align=center] (kg) at (0mm, 75mm) {KG} ;
    \node [draw, fill=gray!10, minimum height=5mm, text depth=.25ex,
      outer sep=0, align=center, right=0mm of kg] (corpus) {Corpus} ;
    \node [marks, above=.1mm of corpus.140] {2} ;
    \node [minimum height=6mm, align=center, above=8mm of qcn2]
    (cand) {Candidate $e_2\in\mathcal{E}_2$} ;
    \node [marks, left=.1mm of cand] {4} ;
    \node [minimum height=6mm, align=center, right=2mm of cand]
    (context) {Contexts~$\{c\}$} ;
    \node [marks, below=.1mm of context.330] {3} ;
    \draw [->] (corpus.south) to (cand) ;
    \draw [->] (corpus.290) to (context.north) ;
    \path [->] (context.south) edge[out=270,in=0] (qcn0.30) ;
    \path [->] (context.south) edge[out=270,in=0] (qcn1.15) ;
    \path [->] (context.south) edge[out=270,in=0] (qcn2.0) ;
    \draw [->] (cand.south) to (qcn0.80) ;
    \draw [->] (cand.south) to (qcn1.85) ;
    \draw [->] (cand.south) to (qcn2.90) ;
    \draw [->] (kg.290) to (cand.north) ;

    \node [align=center, minimum height=6mm, above=8mm of qrn]
    (rel) {Relation~$r$} ;
    \draw [->] (rel) to (qrn.north) ;
    \node [marks, left=.1mm of rel] {8} ;
    \node [marks, left=.1mm of qrn] {9} ;

    \node [align=center, minimum height=6mm, above=8mm of qtn]
    (typ) {Type~$t_2$} ;
    \node [marks, left=.1mm of typ] {6} ;
    \node [marks, left=.1mm of qtn] {7} ;
    \draw [->] (typ) to (qtn.north) ;

    \draw [->] (kg.south) to +(0,-4mm) -| (typ.north) ;
    \draw [->] (kg.250) to +(0,-2mm) -| (rel.north) ;

    \node [minimum height=6mm, minimum width=12mm, inner sep=0mm,
      fill=yellow!10] (query) at (-70mm, 55mm) {Query~$q$} ;
    \draw [->] (query) -| (qrn.120) ;
    \draw [->] (query) -| (qtn.120) ;
    \draw [->] (query) -| (qcn0.120) ;
    \draw [->] (query) -| (qcn1.120) ;
    \draw [->] (query) -| (qcn2.120) ;
    \node [minimum height=6mm, minimum width=12mm, inner sep=0mm,
      fill=yellow!10, below=1mm of query] (e2gold) {Gold~$\mathcal{E}^*_2$} ;

    \node [minimum height=6mm, inner sep=0mm, left=10mm of kg]
    (e1s) {$\mathcal{E}_1=\{e_1\}$} ;
    \draw [->] (e1s) to (kg) ;
    \node [draw, fill=gray!10, minimum height=5mm, minimum width=16mm,
      text depth=.25ex, inner sep=0mm, left=10mm of e1s] (e1link) {$e_1$~Linker} ;
    \node [marks, above=.1mm of e1link] {1} ;
    \draw [->] (e1link) to (e1s) ;
    \draw [->] (query) |- (e1link) ;

    \node [draw, fill=gray!10, align=center,
    	below right=5mm and -30mm of qtn]
    (comb) {Score~combination~network\\ (\S\ref{sec:Combine})} ;
    \node [marks, below=-3mm of comb.340] {10} ;
    \draw [->] (agg.south) |- (comb) ;
    \draw [->] (qtn.south) to (comb.60);
    \draw [->] (qrn.south) to (comb.120);
    
    \node [below=3mm of e2gold, align=center] (loss) {Ranking\\ loss} ;
    \node (rlist) [align=center] at (comb -| loss) {Ranking\\ of $\mathcal{E}_2$};

    \node [marks, right=.1mm of loss] {11} ;
    \node [align=center, left=5mm of comb]
    (score) {Score\\ of~$e_2$} ;
    \draw [->] (comb) to (score) ;
    \draw [->] (score) to (rlist) ;
    \draw [->] (rlist) to (loss) ;
    \draw [->] (e2gold) to (loss) ;
  \end{tikzpicture}
  \caption{AQQUCN block diagram showing $e_1$-linking (1), joint candidate generation from KG and corpus (2), followed by scoring by three convolutional networks for matches with corpus contexts (3--5), target type (6,~7), and relation (8,~9), followed by the score combination network (10) trained with a ranking loss~(11).  Here we show only the query template $e_1 \stackrel{r}{\longbar} e_2$; other cases are similar.  KG+Corpus emit a candidate set $\mathcal{E}_2$ along with multiple interpretations per candidate, but (3--10) are shown as acting on one candidate~$e_2$, to reduce clutter.  After each $e_2$ is scored, $\mathcal{E}_2$ can be ranked.}
\label{fig:aqqucn-block-diagram}
\end{figure}

\subsection{Modifications and new modules}
\label{sec:AqquChanges}

Our implementation of AQQUCN is based on AQQU because it provides a well-written, reusable implementation of query entity linking and interpretation generation.  We modify and enhance AQQU in the following ways, as also illustrated in \figurename~\ref{fig:aqqucn-block-diagram} (for the $e_1 \stackrel{r}{\longbar} e_2$ query template).  The important steps are listed below.  (Item numbers correspond to modules in \figurename~\ref{fig:aqqucn-block-diagram}.)

\begin{table}[th]
\begin{tabular}{|c|l|} \hline
$q$ & Query represented as a string \\
$\mathcal{E}_1$ & Set of grounded entities in $q$ \\
$e_1, e'_1 \in \mathcal{E}_1$ & Specific grounded entities \\
$e_2 \in \mathcal{E}_2$ & Candidate answer entity;
set of candidates \\
$\mathcal{E}^*_2$ & Set of gold (ground truth) answer entities \\
$t_2$ & Target type (given a query interpretation) in KG \\
$r, r'$ & Relation types in KG \\ \hline
\end{tabular}
\caption{Notation summary.}
\label{tab:Notation0}
\end{table}

\begin{enumerate}
\item Given a query $q$, we first identify the set of in-query entities $\mathcal{E}_1$ using the widely used entity tagger TagMe\footnote{As in all QA systems, $e_1$-linking accuracy does affect QA accuracy, but the variation is hard to characterize without a battery of entity linking methods with carefully controlled recall/precision profiles.  AQQU gave slightly better accuracy with TagMe than with its own linker, so we used TagMe for all experiments.  SMAPH~\citep{CornoltiFCRS2014SMAPH} would be a better choice, but it is provided only as a network service, and it needs Google search as yet another level of network service, which has severe usage volume restriction.} \citep{FerraginaS2010TagMe}.
\item The KG neighborhood of each $e_1$ is explored to collect candidate~$e_2$s, similar to prior KBQA (Knowledge Base driven QA) systems \citep{BastH2015aqqu,YihCHG2015STAGG,XuRFHZ2016FbCorpusQA}.  Like AQQU, we limit $\mathcal{E}_2$ to the set of entities occurring in the 2-hop KG neighborhood\footnote{Two hops are needed to traverse mediator nodes like $m$.} of any entity in $\mathcal{E}_1$.
\item As the example in \figurename~\ref{fig:qaCorpusGraph} illustrated, some evidence supporting the correct entity, such as \emph{before} in the query and \emph{prior to} in the snippet, may come from the corpus.  As we shall see in Section~\ref{sec:results}, corpus evidence can greatly augment KG-based evidence.  Therefore, we also gather text snippets (at roughly the granularity of sentences) that mention some $e_1$ or words from the query.  
\item The next step is to identify the set of candidate answer entities, $\mathcal{E}_2$.  Apart from KG neighborhoods of $e_1$s, we also collect (non-$e_1$) entities that occur in any snippet as a candidate~$e_2$.  This allows the system to recover from early errors, such as when $\mathcal{E}_1$ identified by the entity tagger is empty or wrong, or when the query and answer entities are more than two hops apart in the KG.  The union of candidates from KG and corpus are called $e_2$s.
\item We use three neural modules to inform the score combination network, which replaces AQQU's interpretation ranking module.  The query corpus convnet (QCN), described in Section~\ref{sec:qcn}, scores the evidence in the context snippets, given the query and the candidate~$e_2$.
\item Candidates $e_2$ collected from the KG may be accompanied with designated types~$t_2$.
\item The query type convnet (QTN), described in Section~\ref{sec:qtn}, scores the potential presence of a textual clue to $t_2$ somewhere in the query.
\item Candidate $e_2$ found in the KG is also connected to $e_1$ via a relation~$r$.
\item The query relation convnet (QRN), described in Section~\ref{sec:qrn}, scores the potential presence of a textual clue  to $r$ somewhere in the query.
\item The score combination network works on the joint space of candidate interpretations and entities, ending with a ranking of candidate entities that may draw signals from multiple interpretations in general. AQQU scores are used as additional features (not shown).  Outputs from the three convnets are wired together in a markedly non-uniform architecture, consistent with the inference pathways shown in \figurename~\ref{fig:qaCorpusGraph}.  
\end{enumerate}
A summary of notation introduced thus far is given in \tablename~\ref{tab:Notation0}.

Unlike \citet{JoshiSC2014KgCorpusJoint}, AQQUCN does not attempt to segment the query into disjoint spans that describe $e_1, r, t_2$, but lets multiple neural networks run over the query.  This allows AQQUCN to process long queries that \citet{JoshiSC2014KgCorpusJoint} could not.  Moreover, a query span can inform multiple networks; consider queries \query{who discovered penicillin} and \query{who discovered antarctica}, where `who' carries a  lot less information about $t_2$ than the $e_1$ mentions `penicillin' and `antarctica'.

\section{Detailed design of AQQUCN modules}
\label{sec:AqqucnModules}

In this section we present the details of the new modules we added to AQQU:
the query-corpus, query-type and query-relation networks, as well as the score combination network.

\subsection{Query-Corpus Network (QCN)}
\label{sec:qcn}

For each query $q$, we get zero or more evidence snippets from the entity-annotated Web corpus.  Each snippet contains a candidate answer entity $e_2 \in \mathcal{E}_2$.  We use the query corpus network (QCN) for assigning a relevance score to each snippet.   This relevance score, and also the confidence score for linking $e_2$ to a snippet, are features used in the final score combination network of AQQUCN (Section~\ref{sec:Combine}).

Given a query $q$ and a snippet from the Web corpus, QCN should assign high score to the pair if the snippet contains evidence to correctly answer~$q$.  Training data for this network is in the form of positive and negative snippets for each training query. As manual generation of such labels involves considerable effort, we resort to (possibly noisy) indirect supervision instead.  We treat all text snippets centered around a gold answer entity and containing at least one query word (non-stopword) as positively labeled snippets.  Similarly, we treat all text snippets centered around any non-answer entity and containing some or all query words as negatively labeled snippets (examples in Table~\ref{fig:corpusSnippets}).  To train this network, we use the state-of-the-art short text ranking system proposed by \cite{SeverynM2015SiamConvNet}.  Once QCN is trained, we have a score for each snippet belonging to a candidate answer entity $e_2 \in \mathcal{E}_2$.

While Siamese convolutional networks served well in the QCN module, the broad architecture of AQQUCN can accommodate competitive alternatives \citep{LvZ2009positional, PetkovaBC2007NamedEntityProximity, ZhiltsovKN2015FieldedSDM, hui2017pacrr}.  Measuring the effect of this choice on QA accuracy is left for future work.

Match signals from multiple snippets supporting an entity $e_2$ have to be aggregated before passing on to the combination network shown at the bottom of \figurename~\ref{fig:aqqucn-block-diagram}.  Each candidate $e_2$ may have diverse number of supporting snippets.  Usually, a number of standard aggregates (sum, max, etc.) are computed and then a weighted combination learnt \citep{BalogAdR2009LanguageModelExpert, MacdonaldO2006VotingExpert, SawantC2013Fader, JoshiSC2014KgCorpusJoint}.  For our data sets, we found a simple sum of snippet scores to be adequate: we add up the snippet scores over all snippets belonging to an entity $e_2$ for a query $q$, and use it as a feature for ($q$, $e_2$) (feature~1 in Table~\ref{tab:features}).  More complex pooled aggregators can be explored in future work.

\begin{table}[t]
\centering
\begin{tabular}{| p{3cm} | p{8cm} |}
\hline
Query   &  Snippets from Entity-annotated Web corpus \\
\hline
spanish poet died civil war & [Positive] ``\uwave{Lorca} was executed in 1936, during the spanish civil war.'' \\ 
 & [Negative] ``The murder of the spanish poet by \uwave{nationalists} in the civil war remains one of Spain's open wounds.'' \\
\hline
Who was the first U.S. president ever to & [Positive] ``\uwave{Nixon} become the first president in American history to resign.''\\
resign? & [Negative]``\uwave{Gerald R. Ford} took the oath of office after the first-ever resignation by a U.S. President.'' \\
\hline
\end {tabular}
\caption{Example positive and negative corpus snippets for queries. Note how `Lorca' and `Nixon' are mentions of the gold answer entities \protect\path{Federico_Garcia_Lorca} and \protect\path{Richard_Nixon} respectively, while `nationalist' and `Gerald R. Ford' are mentions of non-answer entities \protect\path{Francoist_Spain} and \protect\path{Gerald_Ford} respectively.}
  \label{fig:corpusSnippets}
\end{table}

\begin{table}[t]
\begin{tabular}{| p{40mm} | p{70mm} |}
\hline
Type   &  Type patterns  \\
\hline
\path{/book/author} & dramatist, author, journalist, poet, novelist, writer, editor\\  \hline
\path{/people/deceased_person} & dead, deceased, late, expired, deceased person, victim, person \\  \hline
\path{/film/writer} & screenwriter, writer \\
\hline
\end {tabular}
\caption{Example path patterns for types. Due to automatic extraction, some patterns  may be idiosyncratic (e.g. `victim' for \protect\path{/people/deceased_person} or `injury' for \protect\path{/medicine/medical_treatment}).}
  \label{fig:typeDescription}
\end{table}

\subsection{Query-Type Network (QTN)}
\label{sec:qtn}

The query-type network (QTN) outputs a compatibility score between the query $q$ and a candidate type $t_2$.  This is a multi-class,
multi-label classification problem, as a query may imply more than one correct answer type (e.g. \path{/music/composer} and
\path{/music/artist} for the query \query{saturday night fever music band}).  Good quality training data in the form of (query, type) pairs is essential to ensure that the network learns to handle different types of (or the lack of) query syntax and its correspondence with the answer type.  In our first attempt, we included all $(q, t)$ pairs in the training data, where $t$ was any type connected to the gold answer entity $e_2$ for the query $q$ in the training set.  However, this strategy resulted in many spurious types.  For example, \path{/broadcast/radio_station_owner} is not the correct answer type for the query \query{maya moore college}, even if the answer entity \path{University_of_Connecticut} belongs to that type.  Therefore, we used human supervision.  Paid student volunteers were asked to label $(q, t)$ pairs as correct or incorrect, which helped remove approximately 30\% $(q, t)$ pairs as irrelevant and improve training data quality.


We found that, given the small number of training queries, the data obtained through above process may not be enough to understand the variety of syntax used to imply a type.  For robust training, we resorted to representing the type through additional patterns (Table~\ref{fig:typeDescription}) obtained as follows.
\begin{description}
\item[Freebase relation names:] Consider a (\path{subject},
  \path{relation}, \path{object}) fact triple
  e.g. (\path{Captain_America:_The_First_Avenger},
  \path{/film/film/prequel}, \path{Thor}).  Relations in freebase have
  composite names in the form ``/x/y/z'' where x is a topical domain and 
  y and z indicate the
  types for \path{subject} and \path{object}. E.g. \path{prequel} is a
  type indicator word for \path{Thor}.  Meanwhile, Freebase declares
  expected type for endpoint entities of each $r$. For
  \path{/film/film/prequel}, expected end type is \path{/film/film}. 
  We combine these two information
  nuggets and add \path{prequel} as a pattern expressing the type
  \path{/film/film}.
\item[Freebase type names:] Ending substrings of the type name are
  also considered as patterns (e.g. `treatment' for the type
  \path{medical_treatment}).
\end{description}
At the end of this exercise, we have zero or more patterns for each
type (Table~\ref{fig:typeDescription}).

\begin{figure}[th]
  \centering 
  \begin{tikzpicture}[font={\fontsize{8pt}{10}\selectfont},]
    \matrix(words)[matrix of nodes, column sep=0, row sep=0, inner sep=0,
      outer sep=0, minimum size=5mm, nodes={text width=15mm, align=right}] {
      jimmy\, \\ page\, \\ band\, \\ before\, \\ led\, \\ zeppelin\, \\
    };
    \matrix(redm)[matrix of nodes, right=0 of words-2-1.south east, fill=red!15,
      nodes={draw, anchor=center}, nodes in empty cells,
      column sep=0, row sep=0, inner sep=0, outer sep=0, minimum size=5mm]{
      & & & \\ & & & \\ & & & \\ & & & \\
    };
    \matrix(whtm)[matrix of nodes, below=0 of redm, fill=gray!5,
      nodes={draw}, nodes in empty cells,
      column sep=0, row sep=0, inner sep=0, outer sep=0, minimum size=5mm]{
      & & & \\ & & & \\
    };
    \foreach \j in {1,...,3} {
      \matrix(reda\j)[matrix of nodes,
        below right=\j mm and 9mm-\j *1mm of redm.north east,
        fill=red!15, nodes={draw}, nodes in empty cells,
        column sep=0, row sep=0, inner sep=0, outer sep=0, minimum size=5mm]{ \\ };
      \matrix(whta\j)[matrix of nodes, below=0 of reda\j, fill=gray!5,
        nodes={draw}, nodes in empty cells,
        column sep=0, row sep=0, inner sep=0, outer sep=0, minimum size=5mm]{ \\ \\ };
    }
    \draw [draw=none, fill=gray!50, opacity=0.5] (redm-1-4.north east) -- (reda3.north west)
    -- (reda3.south west) -- (redm-4-4.south east) -- cycle;
    \matrix(poolm)[matrix of nodes,
      right=5mm of whta1, nodes={draw}, nodes in empty cells,
      column sep=0, row sep=0, inner sep=0, outer sep=0, minimum size=5mm]{ \\ \\ \\ };
    \draw [draw=none, fill=gray!50, opacity=0.5] (reda3-1-1.north east)
    -- (poolm-3-1.north west) --  (poolm-3-1.south west) -- (whta3-2-1.south east) -- cycle;
    \matrix(feata)[matrix of nodes,
      below=1mm of poolm, nodes={draw=green!30!black}, nodes in empty cells, fill=green!10,
      column sep=0, row sep=0, inner sep=0, outer sep=0, minimum size=5mm]{ \\ \\ };
    \matrix(outa)[matrix of nodes,
      right=9mm of poolm-3-1, nodes={draw}, nodes in empty cells, fill=orange!10,
      column sep=0, row sep=0, inner sep=0, outer sep=0, minimum size=5mm]{ \\ \\ \\ \\ };
    \draw [draw=none, fill=gray!50, opacity=0.5]
    (poolm.north east)--(outa.north west)--(outa.south west)--(feata.south east)-- cycle;
    \draw node[rotate=90, left=4mm of outa, align=center, anchor=center]
    (sig) {Fully-connected \\ layer with sigmoid};
    \draw node[rotate=90, right=4mm of outa, align=center, anchor=center]
    (score) {Score for each\\ type $t_2$};
    \draw node[left=2mm of feata-2-1.west, align=left, anchor=center, text=green!30!black]
    (overlap) {Features \\ for overlap \\ between \\ query words and $t_2$ };
    \draw node[above=2mm of reda1.north, align=center, anchor=center] (conv) {Convolution};
    \draw node[above=1mm of poolm.north, align=center, anchor=center] (conv) {Pooling};
  \end{tikzpicture}
  \caption{Query-Type Network (QTN) architecture and inputs.  The Query-Relation Network (QRN) architecture is identical except that types are replaced by relations in the training data.}
  \label{fig:typeRelNNArch}
\end{figure}
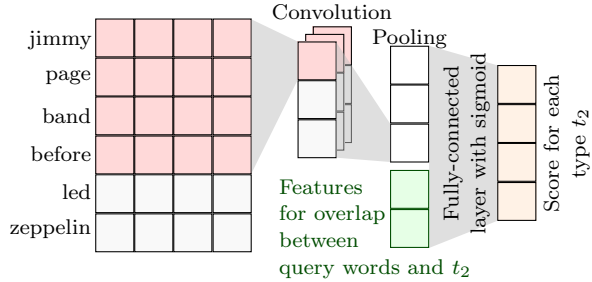

\figurename~\ref{fig:typeRelNNArch} illustrates the design of QTN. This multi-class multi-label architecture is partly inspired
by~\cite{Kim2014ConvnetSentence} and~\cite{SeverynM2015SiamConvNet}. For each query $q$
input to the network, the network provides an output score vector of
size equal to the number of types, as follows:  
\begin{enumerate} 
\item In the initial layer, each query word is represented as a vector embedding learnt during training.  
Then convolution and pooling layers are used to extract a fixed-length feature vector from the variable length input.
\item 
In a separate layer, we compute word overlap features inspired by \cite{SeverynM2015SiamConvNet}.
Specifically, we compute Jaccard similarity between the query 
and each type name, by representing each as a bag of words. 
We also compute Jaccard similarity between the query and each type pattern, 
and then take max over all patterns of a given type. This process results in two features
for each (query, type) pair.
\item Similar to the bag of words based overlap, we compute Jaccard similarity between the word stem (lemmatized) form for each query, type name and type pattern; resulting in two more features for each (query, type) pair.  
\item Once the overlap features as well as convolution-pooling features are computed,
a fully connected hidden layer with sigmoid activation function is
used at the last stage, to score all types.
\end{enumerate}

\subsection{Query-Relation Network (QRN)}
\label{sec:qrn}

The Query-Relation Network (QRN) outputs a compatibility score between a candidate
relation $r$ and the query $q$.  As with QTN, we generate multi-class,
multi-label training data in the form of $(q, r)$ pairs, where $r$ is
a relation connecting to the gold answer entity $e_2$ to the entity
$e_1$ mentioned in the query $q$.  There could be multiple $r$, for
the same ($q, e_1, e_2$) tuple.  Such training data generation process
is common in previous work
\citep{Dong:2015,YihCHG2015STAGG,bordes2014question} and human curation is not
used to remove noise.

\begin{table*}
\centering
\begin{tabular}{| p{6.5cm} | p{4.5cm} |}
\hline
Relation  &  Relation patterns  \\
\hline
\path{/government/government_office_or_title/jurisdiction} & of, president, president of, office \\
\hline
\path{/film/writer/film} & film, film by, by, of, written by, wrote, author of \\  \hline
\path{/theater/play/composer} & by, written by, music by, wrote, with music of, in \\
\hline
\end {tabular}
\caption{Example path patterns for relations. Notice how the same patterns (`written by', `of') can imply different relations, causing ambiguity in query interpretation. }
  \label{fig:relDescription} 
\end{table*}

Similar to QTN, we enrich our training data using relation description
patterns.  To generate these patterns, we start with (\path{subject},
\path{relation}, \path{object}) facts in the KG and locate sentences
in the annotated corpus where both \path{subject} and \path{object}
are mentioned.  We identify the path connecting the two in the
dependency parse of the sentence, expressed as a sequence of
lemmatized words.  We count the number of times each path was found,
and retain only the most frequent paths.  This gives a bag of path
patterns that describe relation $r$ (examples in
Table~\ref{fig:relDescription}).  The network architecture for QRN is same as 
in \figurename~\ref{fig:typeRelNNArch}. The only difference is that for QRN we have
(query, relation) and (corpus pattern, relation) tuples as training instances.

\subsection{Score combination network}
\label{sec:Combine}

Referring back to \figurename~\ref{fig:aqqucn-block-diagram},
QRN, QTN, and multiple QCNs send their scores as features to a final combination network that represents each candidate $e_2$ as a feature vector and scores it in conjunction with\footnote{For simplicity, we describe the single-relation case; multi-hop cases with mediator nodes are handled analogously.} $I = \langle e_1, r, t_2 \rangle$.  In abstract terms, if $I$ denotes an interpretation and $e_2$ a candidate entity, at this point we have a \emph{score matrix} $S(I,e)$ indexed by interpretations as rows and candidate response entities as columns.  

During both training and inference, $I$ is latent.  Standard learning to rank methods \citep{Liu2009LearningToRank} are not directly applicable to our score combination network because of the latent variables implicit in interpretation~$I$.  In fact, the additional complications posed by these latent variables currently limit us to the relatively simple pairwise ranking paradigm with a linear scoring function \citep{Joachims2002ranksvm}.  Direct optimization of listwise and setwise metrics in presence of latent variables is left as future work.  In the rest of this section, we will describe three approaches to train and deploy the score combination network.

\begin{table}[t]
  \centering
  \begin{small}
  \begin{tabular}{|l|l|} 
  \hline
  No. & Description \\
  \hline
1 & Sum of QCN match scores over all snippets for ($q$, $e_2$)\\
 \hline
2--8 & Entity match features 1--7 from~\citet{BastH2015aqqu} \\
 \hline
9 & Sum of QRN match scores over all relations $r$ s.t.\ $(e_1,r,e_2) \in \text{KG}$ \\
 \hline
 10--19 & Relation token match features 8--17 from~\citet{BastH2015aqqu} \\
 \hline
20 & Best QTN match score from all feasible types $t_2$ s.t.\ $(e_1, r, e_2) \in \text{KG}, e_2 \in t_2$ \\
 \hline
 21--26 & General features 18--23 from~\citet{BastH2015aqqu}\\  
 \hline
 27 & AQQU-assigned rank of structured interpretation supporting $e_2$ \\
  \hline
  \end{tabular}
  \end{small}
  \caption{Features used by the score combination network.}
  \label{tab:features}
\end{table}

\subsubsection{Features for score combination}
\label{sec:CombineFeatures}

The score combination module shown at the bottom of \figurename~\ref{fig:aqqucn-block-diagram} uses a feature vector $\phi$ to describe the match between query $q$, each candidate query interpretation $I \in \mathcal{I}$ and each candidate answer entity $e_2 \in \mathcal{E}_2$.   These features are informed by three role-differentiated convolutional networks (described in detail in the rest of this section).  These are
\begin{itemize}
\item the query-corpus network (QCN) described in Section~\ref{sec:qcn},
\item the query-type network (QTN) described in Section~\ref{sec:qtn}, and
\item the query-relation network (QRN) described in Section~\ref{sec:qrn}.
\end{itemize}
As in AQQU, we also include additional features such as entity tagger scores.  Table~\ref{tab:features} shows the complete list of features.

\subsubsection{Single interpretation (AQQUCN-1)}
\label{sec:aqqucn1}

In some data sets like SimpleQuestions \citep{BordesUCW2015MemNetQA}, each query, by construction, can be answered from the KG alone, using exactly one interpretation.  This is largely true of WebQuestions \citep{BerantCFL2013SEMPRE} as well.  As we report later, the post-facto single best (`silver', because the `gold' interpretation is not provided) interpretation retrieves the gold entity set with accuracy much higher than any system.  Therefore, all that remains is to try to infer the silver interpretation.  This is exactly what AQQU attempts to do.  AQQU first aggregates $S(I,e_2)$ over all candidates $e_2$ to get a per-interpretation score, which is used to rank them and choose the top interpretation.  Training is provided by comparing the observed F1 scores of competing candidate interpretations.  Our resulting system, \textbf{AQQUCN-1}, is similar to AQQU, except that we use convnets and draw on corpus information.  The entity set retrieved by the single interpretation $\tilde{I}$ can then be sorted by decreasing~$S(\tilde{I},e)$ for ranking, if needed.

\subsubsection{Allowing a limited number of interpretations (AQQUCN-FEW)}
\label{sec:aqqucnfew}

The assumption that a single interpretation can recall all relevant answer entities may not be valid in all situations.  In particular, as we shall report later, interpretations derived from both KG and corpus can be necessary to cover the gold response entities.

In a set $\mathcal{E}_2$ of candidate entities $e_2$, if the score of each $e_2$ is determined by its best supporting interpretation, then the number of distinct interpretations supporting the candidate set $\mathcal{E}_2$ may approach $|\mathcal{E}_2|$ itself.  In the next section, we will allow that to happen freely.  In this section, we will take a small step to generalize one interpretation to a limited number of interpretations.

Suppose the universe of available interpretations is $\mathcal{I}$, from which we can admit\footnote{We use $K$ for the number of top entities in the response to the user, and $K'$ for the number of interpretations to be used internally.} $\mathcal{I}' \subseteq \mathcal{I}$, with $|\mathcal{I}'| \le K'$, while scoring all the candidate entities.  The score of an entity is thereby restricted from $\max_{I\in\mathcal{I}} S(I,e_2)$ to $S(e_2) = \max_{I\in\mathcal{I}'} S(I,e_2)$.  We are given the set of gold (relevant) entities $\mathcal{E}^+_2$.  Let irrelevant candidates be called~$\mathcal{E}^-_2$.  Then we want $S(e^+_2) > S(e^-_2)$ for any pair $e^+_2\in \mathcal{E}^+_2, e^-_2 \in \mathcal{E}^-_2$, which is turned into a hinge loss $[S(e^-_2) + \Delta - S(e^+_2)]_+$, where $\Delta$ is a margin hyperparameter and $[\bullet]_+ \equiv \max\{0,\bullet\}$ is the hinge or ReLU operator.  Summarizing, the loss we seek to minimize during training is
\begin{align}
\min_{\mathcal{I}' \subseteq \mathcal{I}, |\mathcal{I}'|\le K'}
\sum_{e^+_2 \in \mathcal{E}^+_2} \sum_{e^-_2 \in \mathcal{E}^-_2}
\left[ \max_{I\in\mathcal{I}'} S(I, e^-_2) + \Delta -
\max_{I\in\mathcal{I}'} S(I, e^+_2) \right]_+.  \label{eq:aqqucnfewtrain}
\end{align}
During inference, we do not know $\mathcal{E}^+_2, \mathcal{E}^-_2$.  Therefore we find
\begin{align}
\mathcal{I}^* &=
\argmax_{\mathcal{I}' \subseteq \mathcal{I}, |\mathcal{I}'|\le K'}
\sum_{e_2 \in \mathcal{E}_2}
\max_{I\in\mathcal{I}'} S(I, e_2)  \label{eq:aqqucnfewtest}
\end{align}
and then sort candidate entities $e_2$ by decreasing $\max_{I \in \mathcal{I}^*} S(I,e_2)$.  Both expressions take time to evaluate that are exponential in $K'$, but we expect $K'$ to be very small, usually under~3 (set heuristically).  While we can try to optimize expression \eqref{eq:aqqucnfewtrain} directly to learn model parameters inside $S(I,e_2)$, the objective is highly nonconvex.  We found it better to use the technique in the next section for training and use expression \eqref{eq:aqqucnfewtest} for inference.

\subsubsection{Allowing unlimited supporting interpretations (AQQUCN-ALL)}
\label{sec:lvdt}

If a candidate entity $e_2$ is supported by multiple interpretations $I$, a reasonable view is that the overall score of $e_2$ is $\max_I S(I,e_2)$, from the best supporting interpretation, which induces a ranking among candidate~$e_2$s.  The set of gold $e_2^+$s is then used to define a loss and train the combination network.  We use a pairwise loss, comparing, for a fixed query $q$, a relevant entity $e_2^+$ with an irrelevant entity~$e_2^-$:
\begin{align}
\max_{e_1^+, t_2^+, r^+} \! w \cdot \phi(q, e_1^+, t_2^+, r^+; e_2^+) + 
\xi_{q, e_2^+, e_2^-} &\ge \Delta + \max_{e_1^-, t_2^-, r^-} \!
w \cdot \phi(q, e_1^-, t_2^-, r^-; e_2^-)  \label{eq:PairCons}
\end{align}
where $\Delta$ is a margin hyperparameter.  Note that the best supporting $e_1^+, t_2^+, r^+$ for $e_2^+$ may be different from the best supporting $e_1^-, t_2^-, r^-$ for $e_2^-$.
\begin{itemize}
\item $\mathcal{Q}$ is the set of queries, and $q$ is one query.
\item $e_2^+$ is a relevant entity, $e_2^-$ is an irrelevant entity, for query~$q$.
\item $\phi$ is the feature vector (see Section~\ref{sec:CombineFeatures}) representing an interpretation, composed of $e_1$ (one or more entities mentioned in query $q$), $r$ (relation mentioned or hinted at in $q$), $t_2$ (type mentioned or hinted at in $q$) and $e_2$ (candidate answer entity).  $\phi$ incorporates inputs from the three convnets.
\item $\xi$ is a vector of non-negative slack variables.
\item $C$ a balancing regularization parameter.
\item $w$ is the weight vector to be learnt.
\end{itemize}
The max in the LHS of constraint \eqref{eq:PairCons} leads to nonconvexity, which we address by introducing auxiliary variables $u(q, e_1^+, t_2^+, r^+; e_2^+)$ for each relevant candidate entity in the following optimization.
\begin{align}
\min_{\xi\ge\vec0, w} \quad
& \tfrac{1}{2}\|w\|_2^2 + \frac{C}{|\mathcal{Q}|} \xi \cdot \vec1 
\quad \text{such that} \notag \\
\forall q,e_2^+, e_2^-; e_1^-, t_2^-, r^-: &
  \sum_{e_1^+, t_2^+, r^+} u(q, e_1^+, t_2^+, r^+; e_2^+)\,
  w \cdot \phi(q, e_1^+, t_2^+, r^+; e_2^+)  \notag \\
  & \qquad \ge \Delta - \xi_{q,e_2^+, e_2^-} +
  w \cdot \phi(q, e_1^-, t_2^-, r^-; e_2^-) \label{eq:LVDT} \\
\forall q, e_1^+, t_2^+, r^+; e_2^+: 
& \quad u(q, e_1^+, t_2^+, r^+; e_2^+) \in \{0, 1\} \notag \\
\forall q,e_2^+: & \quad
\sum_{e_1^+, t_2^+, r^+} u(q, e_1^+, t_2^+, r^+; e_2^+) = 1 \notag \\
\forall q,e_2^+,e_2^-: & \quad \xi_{q,e_2^+, e_2^-} \geq 0 \notag
\end{align}
For tractability, we relax the 0/1 constraint over $u$ variables to
the continuous range $[0,1]$:
\begin{align}
\forall q, e_1^+, t_2^+, r^+; e_2^+: 
& \quad u(q, e_1^+, t_2^+, r^+; e_2^+) \in [0, 1]
\end{align}
The relaxation does not correspond to any discrete interpretation, but is a device to make the optimization tractable.  We obtain a local optimum for \eqref{eq:LVDT} by alternately updating $w$ and~$u$.  Each of these is a convex optimization problem.  \figurename~\ref{fig:qiAlgo} shows the pseudocode for inference in our proposed system and can be directly and efficiently solved.  Through optimization \eqref{eq:LVDT}, AQQUCN integrates query interpretation and entity response ranking into a unified framework, rather than a two-stage compile-and-execute strategy common in other QA systems, which effectively gambles on one best structured interpretation.

\begin{figure}[t]
\begin{boxedminipage}{\hsize}
\begin{algorithmic}[1]  \raggedright
\STATE \textbf{input:} query token sequence $q$
\STATE Generate $\mathcal{E}_1$ (query entity set) using entity tagger
\STATE Generate $\mathcal{I}$ = $\{{I}_i = (e_{1i}, t_{2i}, r_i, q) ; e_{1i} \in \mathcal{E}_1 \}$ as potential interpretations, indexed by~$i$
\STATE Generate $\mathcal{E}_2$ = candidate answer entity set reachable from any ${I}_i \in \mathcal{I}$ in KG or corpus
\FORALL{$e_{2j} \in \mathcal{E}_2$}
  \STATE $\emph{bestScore}_j \leftarrow -\infty$
  \FORALL{interpretation ${I}_i$}
    \STATE Generate CNN scores for  $(q, {I}_i, e_{2j})$ using QTN, QCN, QRN
    \STATE Generate other features in Table~\ref{tab:features}
    \STATE Create $\phi_{ij}$, the feature vector for (${I}_i, e_{2j}$) using the above features
    \STATE Score $\phi_{ij}$ using a trained linear model to get $s_{ij}$, the score of (${I}_i, e_{2j}$)
    \STATE $\emph{bestScore}_j \leftarrow \max_i\{ \emph{bestScore}_j, s_{ij} \}$
  \ENDFOR
\ENDFOR
\STATE \textbf{output:} ranking of $e_{2j} \in \mathcal{E}_2$ according to decreasing $\emph{bestScore}_j$
\end{algorithmic}
\end{boxedminipage}
  \caption{High-level pseudocode for AQQUCN-ALL inference.
 Also see \figurename~\ref{fig:aqqucn-block-diagram}.}
  \label{fig:qiAlgo}
\end{figure}

\subsection{End-to-end vs.\ modular training}

In recent years, end-to-end training of complex neural architectures has lost some appeal.  \citet{ShalevshwartzS2016} showed that the sample complexity of end-to-end training can be exponentially larger than the modular training of individual stages.  \citet{Roth2017IjcaiMccarthy} made similar\footnote{Also see Chapter~11 (End-to-end Deep Learning) of \protect\url{http://www.mlyearning.org/}} arguments about some NLP tasks.  With complex notions of `match' (\figurename~\ref{fig:qaCorpusGraph}), a commensurately complex network (\figurename~\ref{fig:aqqucn-block-diagram}) to train, and comparatively few training instances $(q, E_2^*)$ that do not come with gold structured interpretations, we, too, chose modular training of QRN, QTN and QCN, followed by training the score combination network.  It might be argued that the additional labeled data used to train individual modules renders unfair the competition between various systems.  While there is some validity to this protest, many open-domain QA systems already use externally trained word embeddings \citep{BastH2015aqqu, YihCHG2015STAGG, XuRFHZ2016FbCorpusQA}, externally-trained target type recognizers \citep{MurdockKWFFGZK2012TypeCoercion, YavuzGSSY2016AnswerTypeInference}, and augmented training from SimpleQuestions \citep{BordesUCW2015MemNetQA}.

\subsection{Set retrieval vs.\ ranking and the threshold module}
\label{sec:setvsrank}

Choosing $\mathcal{I}'$ and then ranking candidate entities is the most natural method to drive AQQUCN.  In this mode, AQQUCN can be directly compared against any other entity ranking system, such as the one by \citet{JoshiSC2014KgCorpusJoint}.  On the other hand, comparing AQQUCN-FEW and AQQUCN-ALL with other systems that retrieve entity \emph{sets} is not directly possible, unless the desired size of the entity set were  specified, or the ranked list is somehow truncated.  One way to approach this is to threshold the ranked list based on some criterion.  We explore two thresholding strategies.  In the first strategy, we set the score threshold value to $x$\% of the top ranked entity's score (``relative threshold'').  Tuned on held-out data, $x$ turned out to be $0.95$.  In the second strategy (which we refer to as ``ideal threshold'') we threshold at a position which results in the best value of F1 that can be extracted from the ranking output by our system.  As is obvious, the first one provides  unfair advantage to existing KBQA systems, whereas the second provides unfair advantage to our system and merely provides an idealized, non-constructive upper bound on~F1.

\section{Experiments and results}
\label{sec:results}

\subsection{Testbed}
\paragraph{KG:}  We used Freebase \citep{BollackerEPST2008Freebase} as the KG, specifically, the OpenLink Virtuoso snapshot provided with AQQU.  It provides 2.9 billion relation facts on 44 million entities.  The set of answer types, with about 4976 member types, is also curated and provided with AQQU.  It should be possible to adapt other KGs such as WikiData for use with AQQU and AQQUCN.

\paragraph{Annotated corpus:} We used ClueWeb09B \citep{ClueWeb09} as the corpus.  It has about 50 million Web documents in WARC format, same as the rest of ClueWeb09 and ClueWeb12 (over 500 million documents each).  Any corpus in WARC format can be used with AQQUCN, assuming they have useful entity annotations.  For reproducibility, we used the public FACC1 entity annotations released by Google \citep{Google2013FACC1}.  The typical document has 13--15 entities annotated.  Given enough computational capacity, one can run an entity tagger like TagMe \citep{FerraginaS2010TagMe} over the corpus.  The number of tokens in each corpus snippet was limited to 20, based on hyperparamemer tuning in early versions of the system.  We also verified that minor changes to snippet length did not change the results noticeably.

\paragraph{Query sets:}
\citet{JoshiSC2014KgCorpusJoint} provided syntax-poor translations \citep{CsawProject} of syntax-rich queries from TREC and INEX question answering competitions, as well as
a fraction of syntax-rich queries from WebQuestions \citep{BerantCFL2013SEMPRE}.  This gave us four query sets summarized in Table~\ref{fig:querysets} and called TREC-INEX-KW, TREC-INEX, WebQuestions-KW and WebQuestions.

By design, all WebQuestions queries can be answered using the Freebase KG.  In contrast, only 57\% of TREC-INEX queries can be answered from KG alone under the restriction that $e_1$ and $e_2$ lie within two hops.  Thus corpus evidence is important for TREC-INEX.

\begin{table}[t]
  \centering
  \begin{tabular}{|c | c | c | c | c|}
    \hline
    Source & Name & \#train & \#test & Query type\\
    \hline
    TREC and INEX& TREC-INEX-KW   & 493 & 211 & Syntax-poor \\
    \cline{2-5}
    query tracks & TREC-INEX  & 493 & 211 & Syntax-rich \\
    \hline
    WebQuestions& WebQuestions-KW   & 563 & 240 & Syntax-poor \\
    \cline{2-5}
    & WebQuestions  & 3778 & 2032 & Syntax-rich \\
    \hline
  \end {tabular}
\caption{Summary of different query sets.   Syntax-poor queries have also been called `telegraphic' in this paper.  Syntax-rich queries are well-formed natural language questions.  A portion of the train set is internally set aside as the dev(elopment) fold for tuning parameters.}   \label{fig:querysets}
\end{table}

\paragraph{Convnet training protocol:}
Data used to train the convnets is available~\citep{CsawProject}.  Some important design choices are described below.
\begin{description}
\item[QTN and QRN:] Initial word
  vectors are learnt using the CNN-non-static version of
  \cite{Kim2014ConvnetSentence}. Filter sizes are set to 3 and 4 with 150 feature maps each. Drop-out rate is 0.5, with 100 epochs and early stopping
  using a validation split of 10\%. Training is done through
  stochastic gradient descent over shuffled mini-batches with the
  Adadelta update.

\item[QCN:] The width of the convolution filters is set to 5, the
  number of convolutional feature maps 150, batch size 50.
   L2 regularization term is $10^{-5}$ for the parameters of
  convolutional layers and $10^{-4}$ for all the others. The dropout
  rate is set to 0.5. We initialize the word vectors using the word
  embeddings trained by~\cite{Huang:2012}.
\end{description}

\paragraph{Score combination network training protocol:}
All KG paths emanating from a query entity $e_1$ can contribute to
candidate answer set $\mathcal{E}_2$.  We use the pruning process of
\citep{BastH2015aqqu} to restrict the set of KG queries (and hence
$\mathcal{E}_2$) to a practical size.  We normalize all feature values in $[0,1]$ before
sending them to the score combination stage.

\paragraph{Evaluation protocol:}
We evaluate entity ranking using measures common in Information
Retrieval, applied to the response list of entities: mean average
precision (MAP), mean reciprocal rank (MRR), and normalized discounted
cumulative gain at rank 10 (NDCG@10).  For ranking evaluation, the threshold module of Section~\ref{sec:setvsrank} is not applied.  For set retrieval evaluation, the system output entity set $\mathcal{E}_2$ is created by thresholding the ranked list.  It is then compared against gold set $\mathcal{E}^*_2$ to compute recall, precision and~F1.

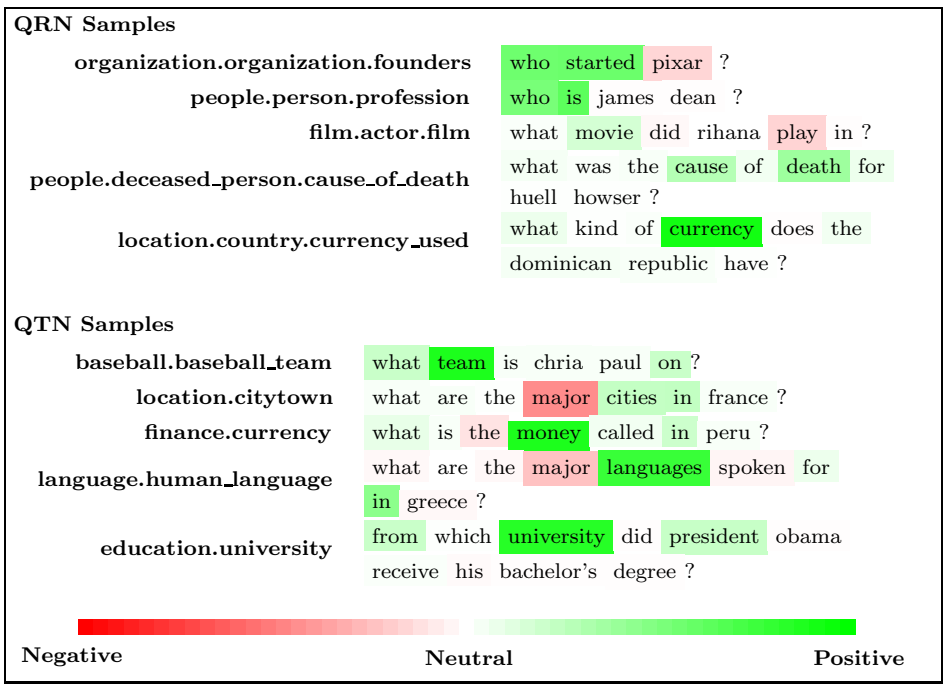
\begin{figure*}[th]
\def\mystrut(#1,#2){\vrule height #1pt depth #2pt width 0pt}    
\begin{fmpage}{0.99\linewidth}
\textbf{QRN Samples}\\
\addvbuffer[3pt 12pt]{\begin{tabular}{rl}
\bf{organization.organization.founders} & \colorbox{green!57}{\mystrut(0,1) who}\colorbox{green!58}{\mystrut(0,1) started}\colorbox{red!16}{\mystrut(0,0) pixar} ? \\

\bf{people.person.profession} &\colorbox{green!53}{\mystrut(0,1) who}\colorbox{green!64}{\mystrut(0,1) is}\colorbox{red!1}{\mystrut(0,0) james}\colorbox{red!1}{\mystrut(0,0) dean} ? \\

\bf{film.actor.film} &\colorbox{green!2}{\mystrut(0,1) what}\colorbox{green!17}{\mystrut(0,1) movie}\colorbox{red!3}{\mystrut(0,0) did}\colorbox{red!0}{\mystrut(0,0) rihana}\colorbox{red!17}{\mystrut(0,0) play}\colorbox{red!1}{\mystrut(0,0) in}? \\

\multirow{ 2}{*}{\bf{people.deceased\_person.cause\_of\_death}} &
\colorbox{green!4}{\mystrut(0,0) what}\colorbox{green!2}{\mystrut(0,0) was}\colorbox{green!3}{\mystrut(0,0) the}\colorbox{green!29}{\mystrut(0,0) cause}\colorbox{green!5}{\mystrut(0,0) of }\colorbox{green!38}{\mystrut(0,0) death}\colorbox{green!7}{\mystrut(0,0) for} \\
&\colorbox{green!1}{\mystrut(0,0) huell}\colorbox{green!1}{\mystrut(0,0) howser}?\\

\multirow{ 2}{*}{\bf{location.country.currency\_used}} &
\colorbox{green!7}{\mystrut(0,0) what}\colorbox{green!1}{\mystrut(0,0) kind}\colorbox{green!2}{\mystrut(0,0) of}\colorbox{green!93}{\mystrut(0,0) currency}\colorbox{red!1}{\mystrut(0,0) does}\colorbox{green!5}{\mystrut(0,0) the} \\
&\colorbox{green!4}{\mystrut(0,0) dominican}\colorbox{green!3}{\mystrut(0,0) republic}\colorbox{green!1}{\mystrut(0,0) have}?\\
\end{tabular}}

\textbf{QTN Samples} \\ 
\addvbuffer[3pt 10pt]{
\begin{tabular}{rl}
\bf{baseball.baseball\_team} & \colorbox{green!21}{\mystrut(0,1) what}\colorbox{green!89}{\mystrut(0,1) team}\colorbox{green!3}{\mystrut(0,0) is}\colorbox{green!3}{\mystrut(0,1) chria}\colorbox{green!3}{\mystrut(0,1) paul}\colorbox{green!21}{\mystrut(0,0) on}? \\

\bf{location.citytown} &\colorbox{green!3}{\mystrut(0,1) what}\colorbox{green!3}{\mystrut(0,1) are}\colorbox{green!1}{\mystrut(0,0) the}\colorbox{red!45}{\mystrut(0,1) major}\colorbox{green!23}{\mystrut(0,1) cities}\colorbox{green!24}{\mystrut(0,0) in}\colorbox{green!3}{\mystrut(0,0) france}?  \\

\bf{finance.currency} &\colorbox{green!9}{\mystrut(0,1) what}\colorbox{green!3}{\mystrut(0,1) is}\colorbox{red!11}{\mystrut(0,0) the}\colorbox{green!89}{\mystrut(0,0) money}\colorbox{green!5}{\mystrut(0,0) called}\colorbox{green!20}{\mystrut(0,0) in}\colorbox{green!3}{\mystrut(0,0) peru}? \\

\multirow{ 2}{*}{\bf{language.human\_language}} &
\colorbox{red!3}{\mystrut(0,0) what}\colorbox{green!0}{\mystrut(0,0) are}\colorbox{red!3}{\mystrut(0,0) the}\colorbox{red!24}{\mystrut(0,0) major}\colorbox{green!78}{\mystrut(0,0) languages}\colorbox{red!4}{\mystrut(0,0) spoken}\colorbox{green!7}{\mystrut(0,0) for} \\
&\colorbox{green!45}{\mystrut(0,0) in}\colorbox{red!3}{\mystrut(0,0) greece}?\\

\multirow{ 2}{*}{\bf{education.university}} &
\colorbox{green!21}{\mystrut(0,0) from}\colorbox{green!1}{\mystrut(0,0) which}\colorbox{green!85}{\mystrut(0,0) university}\colorbox{red!1}{\mystrut(0,0) did}\colorbox{green!22}{\mystrut(0,0) president}\colorbox{red!1}{\mystrut(0,0) obama} \\
&\colorbox{green!2}{\mystrut(0,0) receive}\colorbox{red!2}{\mystrut(0,0) his}\colorbox{red!1}{\mystrut(0,0) bachelor's}\colorbox{red!1}{\mystrut(0,0) degree}?\\\end{tabular}}

\begin{tikzpicture}
\def\basiceval#1{\the\numexpr#1\relax}
\foreach \i in {100,96,92,...,0}
    \fill[red!\i] (\basiceval{50-\i/2} mm,0) rectangle ++(2mm,2mm);
\foreach \i in {0,4,8,...,100}
    \fill[green!\i] (\basiceval{50 + \i/2} mm,0) rectangle ++(2mm,2mm); 
\node[text width=1.5cm] at (0,-0.3) {\bf{Negative}}; 
\node[text width=1.5cm] at (10.4,-0.3) {\bf{Positive}};
\node[text width=1.5cm] at (5.3,-0.3) {\bf{Neutral}};
\end{tikzpicture}
\end{fmpage}
\caption{QRN and QTN heatmaps showing query words that most strongly determine the predicted relations and types.}
\label{fig:heatmaps}
\end{figure*}

\subsection{QRN and QTN heatmaps}

To understand the workings of QTN and QRN, we used the public implementation of 
Local Interpretable Model-Agnostic Explanations (LIME)\footnote{\protect\url{https://github.com/marcotcr/lime}} \citep{lime}. LIME linearly approximates a neural model's behavior around the vicinity of a particular instance to detect the sensitivity of a label decision to input features.  \figurename~\ref{fig:heatmaps} illustrates various sentences, their top predicted classes, and the sensitivity to each query word --- positive, neural, or negative --- in predicting that class.  The observed polarities and intensities were generally intuitive.

\subsection{Entity ranking comparison}

The vast majority of KBQA papers report set retrieval accuracy in terms of recall, precision and F1.  To our knowledge, only \citet{JoshiSC2014KgCorpusJoint} report entity ranking accuracy.  We compare AQQUCN against their system in Table~\ref{fig:finalcomparisonMap}, using standard ranking performance measures: Mean Average Precision (MAP), Mean Reciprocal Rank (MRR) and Normalized Discounted Cumulative Gain (NDCG).  In Table~\ref{fig:finalcomparisonMap}, we see 5--16\% absolute improvement in mean average precision (MAP) over various query sets.  The improvement is statistically significant (at $p < 0.0001$).  Ablation studies in Section~\ref{sec:ablation} suggest some causes for the improvements.

\begin{table}[th]
\centering
\begin{tabular}{| c | l | c | c | c |}
\hline
Data~set    & System       & MAP & MRR & NDCG \\
\hline
\rowcolor{red!5} \cellcolor{white}
TREC-INEX-KW & \citet{JoshiSC2014KgCorpusJoint}
& 0.409  & 0.419  &  0.502\\
\rowcolor{green!5} \cellcolor{white}
& AQQUCN-ALL     & \textbf{0.536}  &  \textbf{0.561}  &  \textbf{0.587}\\
\hline
\hline
\rowcolor{red!5} \cellcolor{white}
TREC-INEX & \citet{JoshiSC2014KgCorpusJoint} 
& 0.358 & 0.362 & 0.426 \\
\rowcolor{green!5} \cellcolor{white}
& AQQUCN-ALL  & \textbf{0.409}   & \textbf{0.420}  &  \textbf{0.445} \\
\hline
\hline
\rowcolor{red!5} \cellcolor{white}
WebQuestions-KW     & \citet{JoshiSC2014KgCorpusJoint}
& 0.377  & 0.401  &  0.474\\
\rowcolor{green!5} \cellcolor{white}
& AQQUCN-ALL      & \textbf{0.525}  &  \textbf{0.543}  &  \textbf{0.575}\\
\hline
\hline
\rowcolor{green!5} \cellcolor{white}
WebQuestions  & AQQUCN-ALL &  \textbf{0.604} &  \textbf{0.615} & \textbf{0.632} \\
\hline
\end {tabular}
\caption{Entity ranking performance comparison with \citet{JoshiSC2014KgCorpusJoint}.  For (relatively long) natural WebQuestions queries, their system could not explore all query segmentations in reasonable time.  (These numbers were obtained without thresholding.)}
\label{fig:finalcomparisonMap}
\end{table}

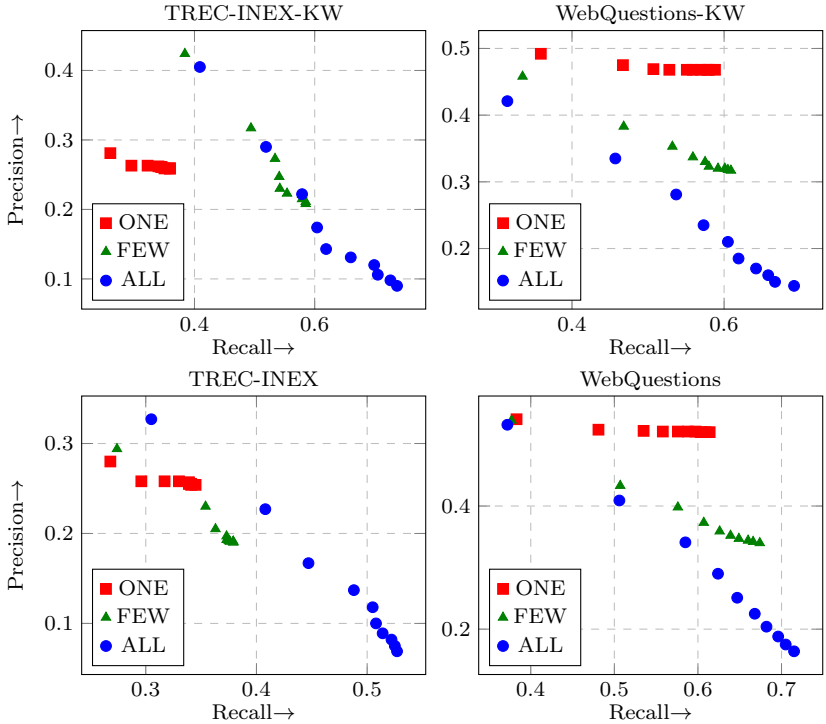
\begin{figure}[ht]
\centering
\begin{tikzpicture}
\begin{axis}[width=.5\hsize,
title style={yshift=-1.5ex,},
ylabel style={yshift=-3ex},
xlabel style={yshift=2ex},
title=TREC-INEX-KW, ylabel=Precision$\rightarrow$,
xlabel=Recall$\rightarrow$,
legend pos=south west, legend columns=1, 
xmajorgrids=true, ymajorgrids=true, grid style=dashed,
scatter/classes={a={mark=square*,red},
b={mark=triangle*,green!50!black},c={blue}}]
\addplot [scatter, only marks, scatter src=explicit symbolic]
table [meta=label] {
x y label
0.26 	0.281 	a
0.295 	0.263 	a
0.322 	0.263 	a
0.34 	0.262 	a
0.345 	0.261 	a
0.345 	0.261 	a
0.35 	0.259 	a
0.359 	0.259 	a
0.359 	0.259 	a
0.359 	0.259 	a
0.384 	0.424 	b
0.494 	0.317 	b
0.534 	0.273 	b
0.541 	0.247 	b
0.542 	0.23 	b
0.554 	0.223 	b
0.579 	0.221 	b
0.579 	0.215	b
0.585 	0.211	b
0.585 	0.208	b
0.409 	0.405 c
0.519 	0.29 c
0.579 	0.222 c
0.604 	0.174 c
0.619 	0.143 c
0.66 	0.131 c
0.699 	0.12 c
0.705 	0.106 c
0.726 	0.098 c
0.737 	0.09 c
} ;
\addlegendentry{ONE}
\addlegendentry{FEW}
\addlegendentry{ALL}
\end{axis}
\end{tikzpicture}
\begin{tikzpicture}
\begin{axis}[width=.5\hsize,
title style={yshift=-1.5ex,},
title=\smash{WebQuestions-KW},
xlabel style={yshift=2ex},
xlabel=Recall$\rightarrow$,
legend pos=south west, legend columns=1, 
xmajorgrids=true, ymajorgrids=true, grid style=dashed,
scatter/classes={a={mark=square*,red},
b={mark=triangle*,green!50!black},c={blue}}]
\addplot [scatter, only marks, scatter src=explicit symbolic]
table [meta=label] {
x y label
0.359 	0.492 	a
0.467 	0.475 	a
0.507 	0.469 	a
0.528 	0.468 	a
0.551 	0.468 	a
0.559 	0.468 	a
0.569 	0.468 	a
0.577 	0.468	a
0.580 	0.468 	a
0.588 	0.468 	a
0.335	0.458	b
0.468	0.383	b
0.532	0.353	b
0.559	0.337	b
0.575	0.330	b
0.580	0.323	b
0.592	0.320	b
0.601	0.320	b
0.605	0.318	b
0.609	0.317	b
0.315	0.421	c
0.457	0.335	c
0.537	0.281	c
0.573	0.235	c
0.605	0.210	c
0.619	0.185	c
0.642	0.170	c
0.658	0.160	c
0.667	0.150	c
0.692	0.144	c
} ;
\addlegendentry{ONE}
\addlegendentry{FEW}
\addlegendentry{ALL}
\end{axis}
\end{tikzpicture}
\begin{tikzpicture}
\begin{axis}[width=.5\hsize,
title style={yshift=-1.5ex,},
ylabel style={yshift=-3ex},
xlabel style={yshift=2ex},
title=TREC-INEX, ylabel=Precision$\rightarrow$,
xlabel=Recall$\rightarrow$,
legend pos=south west, legend columns=1, 
xmajorgrids=true, ymajorgrids=true, grid style=dashed,
scatter/classes={a={mark=square*,red},
b={mark=triangle*,green!50!black},c={blue}}]
\addplot [scatter, only marks, scatter src=explicit symbolic]
table [meta=label] {
x y label
0.268	0.280	a
0.296	0.258	a
0.317	0.258	a
0.330	0.258	a
0.339	0.257	a
0.339	0.256	a
0.339	0.255	a
0.341	0.255	a
0.341	0.254	a
0.345	0.254	a
0.274	0.294	b
0.354	0.230	b
0.363	0.205	b
0.373	0.197	b
0.373	0.193	b
0.375	0.192	b
0.376	0.191	b
0.379	0.191	b
0.379	0.191	b
0.379	0.190	b
0.305	0.327	c
0.408	0.227	c
0.447	0.167	c
0.488	0.137	c
0.505	0.118	c
0.508	0.100	c
0.514	0.089	c
0.522	0.082	c
0.525	0.075	c
0.527	0.069	c
} ;
\addlegendentry{ONE}
\addlegendentry{FEW}
\addlegendentry{ALL}
\end{axis}
\end{tikzpicture}
\begin{tikzpicture}
\begin{axis}[width=.5\hsize,
title style={yshift=-1.5ex,},
title=\smash{WebQuestions},
xlabel style={yshift=2ex},
xlabel=Recall$\rightarrow$,
legend pos=south west, legend columns=1, 
xmajorgrids=true, ymajorgrids=true, grid style=dashed,
scatter/classes={a={mark=square*,red},
b={mark=triangle*,green!50!black},c={blue}}]
\addplot [scatter, only marks, scatter src=explicit symbolic]
table [meta=label] {
x y label
0.383 	0.541 	a
0.481 	0.524 	a
0.535 	0.522 	a
0.558 	0.521 	a
0.576 	0.521 	a
0.588 	0.521 	a
0.597 	0.521 	a
0.603 	0.52 	a
0.609 	0.52 	a
0.614 	0.52 	a
0.377 	0.539 	b
0.507 	0.433 	b
0.576 	0.398 	b
0.607 	0.373 	b
0.626 	0.359 	b
0.639 	0.352 	b
0.649 	0.347 	b
0.66 	0.344 	b
0.666 	0.342 	b
0.674 	0.34 	b
0.372 	0.532 c
0.506 	0.409 c
0.585 	0.341 c
0.624 	0.29 c
0.647 	0.251 c
0.668 	0.225 c
0.682 	0.204 c
0.696 	0.188 c
0.705 	0.175 c
0.715 	0.164 c
} ;
\addlegendentry{ONE}
\addlegendentry{FEW}
\addlegendentry{ALL}
\end{axis}
\end{tikzpicture}
\caption{Interpolated precision vs.\ recall curves for different number of interpretations allowed (without thresholding).}
\label{fig:AQQUCNcomparisonRP}
\end{figure}

\subsection{Effect of number of interpretations allowed}

\figurename~\ref{fig:AQQUCNcomparisonRP} shows interpolated precision against recall, without thresholding, for the variants AQQUCN-1, AQQUCN-FEW and AQQUCN-ALL.  The trends on the two datasets are opposites: on TREC-INEX, performance increases with the number of interpretations, but on WebQuestions, allowing more interpretations reduces F1.  This makes sense, because 85\% of WebQuestions queries can be answered using a single relation \citep{Yao2015LeanQA} and without corpus support.  AQQUCN-FEW restricts the number of interpretations and limits the damage.  

For AQQUCN-1, note that features (2--26) in \tablename~\ref{tab:features} are unique to an interpretation.  Consequently, when an entity has no support from QCN, particularly when the entity is rare or absent in the corpus, the scores of entities retrieved by an interpretation are all the same because of identical feature vectors.  This situation is not so rare that we can ignore its effects.  In such cases, even if AQQUCN-1 retrieves a set of reasonable quality, its ranking results from arbitrary tie-breaking.  AQQUCN-ALL and AQQUCN-FEW are largely immune to this problem, because ties are less likely among the entity scores $\max_I S(I,e_2)$.


Table~\ref{fig:AQQUCNcomparisonF1} reports F1 scores for AQQUCN-1, AQQUCN-FEW and AQQUCN-ALL, after applying thresholding.  The trends are similar to those shown in \figurename~\ref{fig:AQQUCNcomparisonRP}. WebQuestions has somewhat larger $\mathcal{E}^*_2$ on average (natural queries: 2.4; telegraphic: 2.1), but KG-based interpretations are frequently adequate.  Each KG-based interpretation covers more entities, and therefore, relatively fewer interpretations are needed to cover $\mathcal{E}^*_2$.  In contrast, TREC-INEX has smaller $\mathcal{E}^*_2$s on average (natural queries: 1.5; telegraphic: 1.4).  But TREC-INEX depends on corpus-based interpretations, each of which typically covers only one entity.  Therefore, TREC-INEX needs \emph{more} interpretations to get good F1 scores.

\begin{table}[th]
\centering
\begin{tabular}{|l|c|c|c|} \hline
$\downarrow$Data~set & AQQUCN-1 & AQQUCN-FEW & AQQUCN-ALL \\
\hline
TREC-INEX-KW 
& \cellcolor{red!8}{0.264}
& \cellcolor{yellow!10}{0.388}
& \cellcolor{green!24}{0.417} \\
TREC-INEX
& \cellcolor{red!8}{0.269}
& \cellcolor{yellow!10}{0.285}
& \cellcolor{green!24}{0.323} \\ \hline
WebQuestions-KW
& \cellcolor{green!24}{0.492}
& \cellcolor{yellow!5}{0.437}
& \cellcolor{red!4}{0.392} \\
WebQuestions
& \cellcolor{green!24}{0.532}
& \cellcolor{yellow!5}{0.512}
& \cellcolor{red!4}{0.497} \\ \hline
\end{tabular}
\caption{F1 (after thresholding) comparison with different number of distinct interpretations.   $K'=2$ was used for AQQUCN-FEW.  The {\bf best} performing variant of AQQUCN will be referred to as {\bf AQQUCN-Best} hereafter.}
\label{fig:AQQUCNcomparisonF1}
\end{table}

Given the almost exclusive focus on WebQuestions and SimpleQuestions in prior work, these important considerations were discovered only after instrumenting AQQUCN.  Hereafter, we will refer to the best performing variant among AQQUCN-1, AQQUCN-FEW and AQQUCN-ALL as AQQUCN-Best.

\begin{table}[th]
\centering
\begin{tabular}{| c | l | c |c|c|}
\hline
Data  & System  & F1 \\ 
\hline
TREC-INEX-KW
& \citet{BerantL2015AgendaIL} & 0.127 \\
& AQQU \citep{BastH2015aqqu} & 0.222 \\ 
& AQQUCN-Best (AQQUCN-ALL) & \cellcolor{green!30}0.417 \\
& AQQUCN-ALL (ideal threshold) & \cellcolor{blue!10} 0.578 \\ 
& KG+Corpus best single interpretation & \cellcolor{blue!10} 0.362\\
\hline
\hline
TREC-INEX
& \citet{BerantL2015AgendaIL} & 0.107  \\
& AQQU \citep{BastH2015aqqu} & 0.258 \\ 
& AQQUCN-Best (AQQUCN-ALL) & \cellcolor{green!30} 0.323 \\
& AQQUCN-ALL (ideal threshold) & \cellcolor{blue!10} 0.435 \\ 
& KG+Corpus best single interpretation & \cellcolor{blue!10} 0.395 \\
\hline
\hline
WebQuestions-KW
& \citet{BerantL2015AgendaIL} & 0.365\\
& AQQU \citep{BastH2015aqqu}  & 0.470 \\
& AQQUCN-Best (AQQUCN-1) & \cellcolor{green!30} 0.492 \\
& AQQUCN-ALL (ideal threshold) & \cellcolor{blue!10} 0.570 \\ 
& KG+Corpus best single interpretation & \cellcolor{blue!10} 0.698 \\
\hline
\hline
WebQuestions
& \citet{YaoVD2014Jacana} & 0.330\\ 
& \citet{BerantCFL2013SEMPRE} & 0.357 \\
& \citet{Yao2015LeanQA} & 0.443\\ 
& \citet{BerantL2015AgendaIL} & 0.496 \\
& AQQU \citep{BastH2015aqqu} & 0.521 \\ 
& STAGG \citep{YihCHG2015STAGG} & 0.525 \\
& Text2KB \citep{SavenkovA2016KbCorpusQa} & 0.525\\
& Text2KB+STAGG & 0.532 \\
& AQQUCN-Best (AQQUCN-1) & \cellcolor{yellow!20} 0.532 \\
& \citet{XuRFHZ2016FbCorpusQA} & \cellcolor{green!10} 0.533 \\
& Text2KB+STAGG (ideal threshold) & \cellcolor{blue!7} 0.606\\
& AQQUCN-ALL (ideal threshold) & \cellcolor{blue!10} 0.634 \\
& KG+Corpus best single interpretation & \cellcolor{blue!10} 0.737 \\
\hline
\end {tabular}
\caption{F1 comparison with recent KGQA systems.  See text
  for a discussion on relative vs.\ ideal threshold. Note that the results reported for AQQU~\citep{BastH2015aqqu} are with our own runs of their code and are better than the results reported by them, presumably because we used a different entity tagger \citep{FerraginaS2010TagMe}.}
\label{fig:finalcomparisonF1}
\end{table}

\subsection{Entity set retrieval comparison}
\label{sec:rankingToSet}

In \tablename~\ref{fig:finalcomparisonF1}, we compare F1 of entity set retrieval across several\footnote{For three cases, only AQQU \citep{BastH2015aqqu} and Sempre \citep{BerantL2015AgendaIL} code were available.  Text2KB is available at \protect\url{https://github.com/DenXX/aqqu}, but with missing corpus files and no format specification.} KBQA systems \citep{CodaLabWebQuestions}.  AQQUCN is presented with both relative threshold (AQQUCN-Best) and ideal threshold, to separate the quality of ranking vs.\ thresholding.  Also quoted is the F1 score achievable in principle if the single best interpretation is used from KG and corpus.

The first striking observation on \tablename~\ref{fig:finalcomparisonF1} is that, for TREC-INEX, using a single interpretation is a terrible plan; both AQQUCN-Best and AQQUCN-ALL with ideal threshold are far better.  Predictably, without access to a corpus, \citet{BerantL2015AgendaIL} and \citet{BastH2015aqqu} perform poorly.  In sharp contrast, owing to the nature of WebQuestions, a clairvoyant choice of the single best interpretation beats everything else (including \citet{SavenkovA2016KbCorpusQa} with ideal threshold) by a very wide margin.  While AQQUCN with ideal threshold far exceeds all other systems, it is not even close to the best single interpretation.  In terms of non-clairvoyant achievable accuracies, AQQUCN-Best is visibly best for WebQuestions-KW, and ranked second for WebQuestions.  Clearly a great deal of ground has been covered since 2013 for WebQuestions, yet there is plenty of room to improve.  It is also clear that AQQUCN is much better at ranking than thresholding.

\begin{table}[ht]
\centering
\begin{tabular}{| c | c | c | c | c |}
\hline
Data    & System       & F1 \\
\hline
      & AQQUCN-Best  & \cellcolor{green!25}{0.417} \\
      & No QCN  & \cellcolor{red!20}{0.167} \\
TREC-INEX-KW        & No QTN & \cellcolor{red!5}0.410 \\
       & No QRN &  \cellcolor{red!5}0.412 \\
\hline
\hline
      & AQQUCN-Best  & \cellcolor{green!25}{0.323} \\
       &No QCN  & \cellcolor{red!18}{0.192}\\
TREC-INEX       & No QTN & \cellcolor{red!5}0.317 \\
       & No QRN &  \cellcolor{red!5}0.320 \\
\hline
\hline
      & AQQUCN-Best  & \cellcolor{green!25}{0.492} \\
		& No QCN  & \cellcolor{red!5}{0.480}  \\
 WebQuestions-KW        & No QTN & \cellcolor{red!20}0.475 \\
       & No QRN & \cellcolor{red!5}0.489  \\
\hline
\hline
      & AQQUCN-Best  & \cellcolor{green!25}0.532 \\
	& No QCN & \cellcolor{red!5}0.526  \\
WebQuestions          & No QTN & \cellcolor{red!5}0.527 \\
       & No QRN & \cellcolor{red!5}0.529 \\
\hline
\end {tabular}
\caption{Ablation shows the relative effectiveness of the QCN, QTN and QRN networks.}
  \label{fig:convnetBenefits}
\end{table}

\subsection{Ablation tests} \label{sec:corpusevidence}
\label{sec:ablation}

To understand the contributions of the three convnets to the score combination network, we removed each network in turn, re-trained the best performing model, and tabulated the resulting F1 scores in \tablename~\ref{fig:convnetBenefits}.  TREC-INEX (in both query forms) suffers a serious hit if QCN is removed.  This makes sense because of the critical evidence brought in by corpus snippets in case of TREC-INEX.  The effect of removing QCN is smaller for WebQuestions, but still visible, showing that, even if gold entities are in the KG, corpus evidence can help score them better.

The influence between QRN and QTN is more nuanced.  The relation $r$ involved in a query may asserts strong selectional preferences on the types of participating entities.  Therefore, an accurate $r$ predicted by the QRN can often make up for mistakes made in predicting $t_2$ by the QTN.  Conversely, accurately predicting $t_2$ may mitigate a misleading choice of~$r$.  Overall, though, \tablename~\ref{fig:convnetBenefits} shows at least some performance reduction if either QRN or QTN is removed.  For the terse queries in WebQuestions-KW, removing QTN hurts more than removing QCN.

\subsection{Wins and losses}
\label{sec:Discussion}

We performed a side-by-side analysis of a sample of queries for which we found  our system to be doing better and worse than related work. Our system improved on some queries containing qualifiers such as `first', `oldest', since we harness signals from the text corpus. For example, \query{Who was the first U.S. president ever to resign?} can be translated to a complex graph query involving a max/sort over dates, making it difficult to interpret it using only the knowledge graph.  \cite{YihCHG2015STAGG} handled some of these queries using extensively hand-engineered features.  However, such information was readily found in the Web corpus (examples in \figurename~\ref{fig:corpusSnippets}). The corpus also helped when the KG was incomplete (e.g., \query{president sworn on airplane}) or for answering queries with no clear $e_1$ (e.g., \query{which kennedy died first?}).

We performed worse on some queries, especially when the corpus signal added more noise than information and our type or relation CNNs were not able to narrow down to the correct answer. Some of these queries had a non-trivial syntactic structure, possibly not captured by the corpus. For example, \query{What nation is home to the Kaaba?}.  At times, high annotation density in the corpus promoted popular non-answer entities over not-so-popular answer entities.  For the query \query{creator of the daily show}, \path{Jon_Stewart} ranked above \path{Madeleine_Smithberg}, purely based on corpus popularity.  Such cases highlight an opportunity to improve our corpus, type and relation CNNs.


\section{Conclusion}
\label{sec:End}

We presented AQQUCN, a system that unifies structured interpretation of queries with ranking of response entities.  Apart from seamlessly integrating corpus and KG information, AQQUCN has two salient features: it can deal with the full spectrum of query styles between keyword queries and well-formed questions; and it directly ranks response entities, rather than `compile' the input to a structured query and execute that on the KG alone.

\smallskip
\paragraph*{Acknowledgment:}
Thanks to the reviewers for their constructive suggestions. Thanks to Elmar Hau{\ss}mann for generous help with AQQU. Thanks to Doug Oard for advice on set vs.\ ranked retrieval. Thanks to Saurabh Sarda for migrating the code of \citet{JoshiSC2014KgCorpusJoint} to use AQQU.  Partly supported by grants from IBM and nVidia.

\bibliographystyle{spbasic}
\begin{small}
\bibliography{voila,uma}  
\end{small}

\end{document}